\documentclass[
10pt, % Main document font size
a4paper, % Paper type, use 'letterpaper' for US Letter paper
oneside, % One page layout (no page indentation)
headinclude,footinclude, % Extra spacing for the header and footer
BCOR5mm, % Binding correction
]{scrartcl}

\usepackage[
nochapters, % Turn off chapters since this is an article        
beramono, % Use the Bera Mono font for monospaced text (\texttt)
eulermath,% Use the Euler font for mathematics
pdfspacing, % Makes use of pdftex’ letter spacing capabilities via the microtype package
dottedtoc % Dotted lines leading to the page numbers in the table of contents
]{classicthesis} % The layout is based on the Classic Thesis style

\usepackage{arsclassica} % Modifies the Classic Thesis package

\usepackage[T1]{fontenc} % Use 8-bit encoding that has 256 glyphs

\usepackage[utf8]{inputenc} % Required for including letters with accents

\usepackage{graphicx} % Required for including images
\graphicspath{{Figures/}} % Set the default folder for images

\usepackage{enumitem} % Required for manipulating the whitespace between and within lists

\usepackage{lipsum} % Used for inserting dummy 'Lorem ipsum' text into the template

\usepackage{subfig} % Required for creating figures with multiple parts (subfigures)

\usepackage{amsmath,amssymb,amsthm} % For including math equations, theorems, symbols, etc

\usepackage{varioref} % More descriptive referencing

\theoremstyle{definition} % Define theorem styles here based on the definition style (used for definitions and examples)

\theoremstyle{plain} % Define theorem styles here based on the plain style (used for theorems, lemmas, propositions)

\theoremstyle{remark} % Define theorem styles here based on the remark style (used for remarks and notes)

\hypersetup{
colorlinks=true, breaklinks=true, bookmarks=true,bookmarksnumbered,
urlcolor=webbrown, linkcolor=RoyalBlue, citecolor=webgreen, % Link colors
pdftitle={}, % PDF title
pdfauthor={\textcopyright}, % PDF Author
pdfsubject={}, % PDF Subject
pdfkeywords={}, % PDF Keywords
pdfcreator={pdfLaTeX}, % PDF Creator
pdfproducer={LaTeX with hyperref and ClassicThesis} % PDF producer
}
\title{\normalfont\spacedallcaps{Gradient and GENERIC  time evolution towards  reduced dynamics}} % The article title

\author{\spacedlowsmallcaps{Miroslav Grmela*\textsuperscript{1} \& V{\' a}clav Klika\textsuperscript{2} \& Michal Pavelka\textsuperscript{3}}}

\date{} % An optional date to appear under the author(s)

\newcommand{\uu}{{\boldmath \mbox{$u$}}}

\newcommand{\NNN}{{\mbox{$\mathcal{N}$}}}
\newcommand{\MMM}{{\mbox{$\mathcal{M}$}}}
\newcommand{\VVV}{{\mbox{$\mathcal{V}$}}}

\newcommand{\JJJ}{{\mbox{$\mathcal{J}$}}}
\newcommand{\KKK}{{\mbox{$\mathcal{K}$}}}

\newcommand{\rr}{{\boldmath \mbox{$r$}}}

\newcommand{\pp}{{\boldmath \mbox{$p$}}}
\newcommand{\vv}{{\boldmath \mbox{$v$}}}

\begin{document}

%----------------------------------------------------------------------------------------
%	HEADERS
%----------------------------------------------------------------------------------------

%\renewcommand{\sectionmark}[1]{\markright{\spacedlowsmallcaps{#1}}} % The header for all pages (oneside) or for even pages (twoside)
%\renewcommand{\subsectionmark}[1]{\markright{\thesubsection~#1}} % Uncomment when using the twoside option - this modifies the header on odd pages
\lehead{\mbox{\llap{\small\thepage\kern1em\color{halfgray} \vline}\color{halfgray}\hspace{0.5em}\rightmark\hfil}} % The header style

\pagestyle{scrheadings} % Enable the headers specified in this block

%----------------------------------------------------------------------------------------
%	TABLE OF CONTENTS & LISTS OF FIGURES AND TABLES
%----------------------------------------------------------------------------------------

\maketitle % Print the title/author/date block

\setcounter{tocdepth}{2} % Set the depth of the table of contents to show sections and subsections only

\tableofcontents % Print the table of contents

\listoffigures % Print the list of figures

\listoftables % Print the list of tables

%----------------------------------------------------------------------------------------
%	ABSTRACT
%----------------------------------------------------------------------------------------

\section*{Abstract} % This section will not appear in the table of contents due to the star (\section*)
Let $(\MMM,\JJJ)$ be a dynamical model of macroscopic systems and $(\NNN,\KKK)$ a less microscopic model (i.e. a model involving less details) of the same macroscopic systems; $\MMM$ and $\NNN$ are manifolds, $\JJJ$ are vector fields on $\MMM$, and $\KKK$ are vector fields on $\NNN$.
Let $\mathcal{P}$ be the phase portrait corresponding to $(\MMM,\JJJ)$ (i.e. $\mathcal{P}$ is  the set of all trajectories in $\MMM$ generated by a family of vector fields in  $\JJJ$), and $\mathcal{R}$ the phase portrait corresponding to $(\NNN,\KKK)$.  Thermodynamics in its general sense is a pattern recognition process in which $\mathcal{R}$ is recognized as a pattern in $\mathcal{P}$.
In particular, the classical (both equilibrium and nonequilibrium)  thermodynamics arises in the investigation of relations between models $(\MMM,\JJJ)$ and models without time evolution, i.e. models with  $\KKK\equiv 0$. In such case $\mathcal{R}$ is a submanifold of $\MMM$ composed of fixed points. Let $S^{\uparrow}: \MMM\rightarrow \mathbb{R}$ be a potential, called an upper entropy,  generating the vector field $J$.  The equilibrium thermodynamic relation  in $\NNN$ is the lower entropy $S^{\downarrow}(y)$ defined by  $S^{\downarrow}(y)=S^{\uparrow}(x)|_{x=y}$, where $x\in \MMM$, $\NNN\ni y=\hat{x}$, and $\hat{x}$ is a final destination (i.e. when the time $\rightarrow\infty$) of $x$ in the time evolution generated by the vector field $J$.
In this paper we show that if $\KKK \neq 0$ (e.g. in externally forced or, in other words, open systems) then the reduction also provides thermodynamics (we call it flux-thermodynamics). %In this paper we show that if $\KKK\neq 0$ then the reduction also provides thermodynamics (we call it flux-thermodynamics).
If certain conditions are satisfied, then the lower entropy $\mathfrak{S}^{\downarrow}$, that arises in the investigation of the approach $\JJJ\rightarrow \KKK$,%\todo{neni tu lepsi $\JJJ\rightarrow \KKK$?}
is the time derivative of the lower entropy $S^{\downarrow}$ arising in the investigation of the  approach $\MMM\rightarrow \NNN$ as $t\rightarrow \infty$.

%----------------------------------------------------------------------------------------
%	AUTHOR AFFILIATIONS
%----------------------------------------------------------------------------------------

\let\thefootnote\relax\footnotetext{* \textit{email: miroslav.grmela@polymtl.ca}}
\let\thefootnote\relax\footnotetext{1 \textit{\'{E}cole Polytechnique de Montr\'{e}al, C.P.6079 suc. Centre-ville,  Montr\'{e}al, H3C 3A7,  Qu\'{e}bec, Canada}}
\let\thefootnote\relax\footnotetext{1 \textit{Czech Technical University in Prague, FNSPE, Department of Mathematics, Trojanova 13, 120 00 Prague, Czech Republic}}
\let\thefootnote\relax\footnotetext{1 \textit{Mathematical Institute, Faculty of Mathematics, Charles University, Sokolovsk\'{a} 83, 18675 Prague, Czech Republic}}

%----------------------------------------------------------------------------------------

\newpage % Start the article content on the second page, remove this if you have a longer abstract that goes onto the second page

%----------------------------------------------------------------------------------------
%	INTRODUCTION
%----------------------------------------------------------------------------------------

\section{Introduction}

%\todo{Propojit se spec issue - nejake instrukce ci naznaceni od Petera Vane?}
%\todo{(Transitivita - odsouvame; neni to tu nutne)}
%\todo{Souvislost s clankem o produkci entropie a jeji roli pri prechodu mezi levelama; souvislost s clankem o DynMaxEtu z 2019}
%\todo{Myslim, ze se v textu, zejmena ke konci, role konjugovanych promennych vzhledem k entropii zamenuje s konjug prom vzhl k energii.. Chtelo by to sjednotit/opravit.}

Behavior of macroscopic physical systems can be observed and  modeled on different levels of description. The levels differ in the amount of details taken into account in the observations and in  the modeling. The levels with more details are called more microscopic or equivalently less macroscopic. For example the level of kinetic theory (i.e. the level on which one particle distribution function serves as the state variable) is more microscopic than the level of fluid mechanics (where the hydrodynamic fields serve as the state variables).
Let $(\MMM,\JJJ)$ and $(\NNN,\KKK)$ be two models of the same collection of macroscopic systems, the former is more microscopic than the latter. By $\MMM$ we denote  the state space ($\MMM$ is a manifold) of the  model $(\MMM,\JJJ)$, $\JJJ$  is the set of vector fields on $\MMM$. Similarly, the manifold $\NNN$ is the state space of the more macroscopic model and  $\KKK$ is the set of vector fields on $\NNN$.  In order to bring our terminology as close as possible to the terminology used in thermodynamics we use $\JJJ$ instead of
$\mathfrak{X}(\MMM)$ that is the  symbol used in geometry to denote vector fields on $\MMM$. The elements of $\MMM$ are denoted by the symbol $x$, elements of $\JJJ$, called fluxes, are denoted by the symbol $J$. We  use the term ``flux'' in a more general sense than it is used in the context of the local-conservation-law time evolution equations. In this paper a flux $J$ denotes  the complete  right hand side of the equation governing the  time evolution of $x$. Similarly, $\KKK\equiv \mathfrak{X}(\NNN)$, $y\in \NNN$, and $K\in \KKK$.

We introduce  moreover the  notion of  phase portrait. The phase portrait  $\mathcal{P}$ is a collection of all trajectories in $\MMM$ generated by a family of  vector fields $J\in\JJJ$. Similarly, we define the phase portrait $\mathcal{R}$ for the model $(\NNN,\KKK)$

\textbf{\textit{Thermodynamics in a general sense is a theory of relations among  mesoscopic dynamical models   of macroscopic systems. Models involving less details  are related to (are reduced from) models involving more details. The reduction process is a pattern recognition process in which the phase portrait $\mathcal{R}$ of the reduced model is recognized as a pattern in the phase portrait $\mathcal{P}$ of the model involving more details.}}

Thermodynamics is thus a meta-physics since it is a theory of theories. Direct experimental observations are made separately on both levels $(\MMM,\JJJ)$ and $(\NNN,\KKK)$. The experimental evidence for the general thermodynamics is only indirect. It is  obtained by comparing the experimental observations on both levels. An important exception is when $(\MMM,\JJJ)$ is the completely microscopic theory $(\MMM^{(micro)},\JJJ^{(micro)})$ in which macroscopic systems are seen as  composed of $\sim 10^{23}$ atoms and the less detailed model $(\NNN,\KKK)$ is  the classical equilibrium thermodynamics, i.e.  $(\NNN,\KKK)=(\NNN^{(eq)},\KKK^{(eq)})$;  $(E,V,N)\in \NNN^{(eq)}$ and $\KKK^{(eq)}\equiv 0$ (i.e. there is no time evolution taking place on the level of the classical equilibrium thermodynamics). By  $E$ we denote  the internal energy, $V$ is the volume of the macroscopic systems, and $N$ is the number of moles. In this case the ready availability of thermodynamic walls (that either allow to pass freely or block the passage%\todo{vk: je tato veta spravne? co treba 'allow free pass of or block the ...'}  
of the internal energy $E$)  makes a direct experimental access to the entropy $S$ (by measuring the temperature that is  the inverse of the derivative of  $S$ with respect to $E$) that, as we shall see later in this paper, addresses the relation between $(\MMM^{(micro)},\JJJ^{(micro)})$ and $(\NNN^{(eq)},0)$. This then makes  the classical equilibrium thermodynamics a very practically important theory combining microscopic mechanics with heat. Moreover, our innate ability to sense the temperature is essential for our survival since the chemical reactions that take  place inside our bodies and  drive  our actions depend strongly on the temperature.

In this paper we investigate relations among   models $(\MMM,\JJJ)$ and $(\NNN,\KKK)$ that both  involve the time evolution.
We show that by making the reduction (i.e. by recognizing $\mathcal{R}$ as a pattern in $\mathcal{P}$)  we do not only express the vector fields $\KKK$ in terms of the more microscopic vector fields $\JJJ$ (in other words, we do not only provide a microscopic specification of the constitutive relations \footnote{We use the term ``constitutive relations'', the term introduced in the context of the local-conservation-law equations, to denote a specification of $K\in\KKK$} in the model $(\NNN,\KKK)$)  but also bring something completely new into the model $(\NNN,\KKK)$, something that is absent in $(\NNN,\KKK)$ if it is considered only as an autonomous mesoscopic model based only on its own experimental basis. The new addition to $(\NNN,\KKK)$ that arises in the reduction is the fundamental thermodynamic relation for the model $(\NNN,\KKK)$. We call it a fundamental flux-thermodynamic relation. In the mathematical formulation it is a geometry of $(\NNN,\KKK)$ obtained by seeing it as a submanifold inside $(\MMM,\JJJ)$. The mesoscopic model $(\NNN,\KKK)$ becomes enriched by properties that are inherited from a more microscopic viewpoint of the macroscopic systems under consideration.

Model $(\MMM,\JJJ)$ with which we begin our investigation are assumed to possess the Hamiltonian structure. The analysis is illustrated on the particular example in which $(\MMM,\JJJ)$ represents the kinetic theory  and $(\NNN,\KKK)$ is either the classical equilibrium thermodynamics $(\NNN^{(eq)},0)$ or the level of fluid mechanics. If we ignore details of the time evolution leading from $(\MMM,\JJJ)$ to $(\NNN,\KKK)$ then the mapping $(\MMM,\JJJ)\mapsto (\NNN,\KKK)$ appears to be a reducing Legendre transformation expressing the Maximum Entropy Principle (MaxEnt principle). The relation between $(\MMM,\JJJ)$ and $(\NNN,\KKK)$ is presented in this way in Section \ref{2}. The entropy, its maximization, and the reducing projections, that constitute the input of the MaxEnt principle, then emerge   in Section \ref{3} in the investigation of the time evolution in $\MMM$ generated by the vector field $\JJJ$.

\section{Static Reductions}\label{2}

The reduction is presented in this section as a formal mathematical transformation known as the MaxEnt principle.  The transformation is formal, since it lacks a physical justification.%\todo{Myslim, ze jsme psali v DynMaxEntu (a asi i v knizce), ze trochu vyznam ma - least biased estimate}. 
The advantage of the MaxEnt formulation of the reduction is its clarity and possible interpretation based on information theory \cite{jaynes,pkg,jizba-maxent}.%\todo{citace na RedExt, knihu, DynMaxEnt}. 
Its physical basis, that lies in a complex pattern-recognition type analysis of the time evolution that takes place in the initial $(\MMM,\JJJ)$ model, is discussed in the following sections, see also \cite{redext,dynmaxent,pkg}. We present the MaxEnt formulation first (in Section \ref{2.1}) for the reduction towards the equilibrium level $(\NNN^{(eq)},0)$  and then in Section \ref{2.2} to  a mesoscopic level $(\NNN,\KKK)$ involving time evolution.

\subsection{Static reductions $(\MMM,\JJJ)\rightarrow (\NNN^{(eq)},0)$}\label{2.1}

We  discuss separately two examples of $(\MMM,\JJJ)$.

\subsubsection{$y^{(eq)}\mapsto y^{(eq)}$}

In this example we take $(\MMM,\JJJ)\equiv (\NNN^{(eq)},0)$.
We choose $\NNN^{(eq)}\ni y^{(eq)}=(E,N)$, where $E$ is the internal energy and $N$ is the number of moles. We are omitting the volume $V$ since we consider the volume of the region of  $\mathbb{R}^3$ in which the macroscopic systems under consideration are confined as a constant that remains unchanged. No boundary  effects are considered. The extensive quantities (i.e. the quantities that depend on the volume) are assumed to be homogeneous functions of the volume of degree one. Consequently, we put the volume $V$ equal to one and $V$ thus completely disappears from our analysis.

The point of departure  of the transformation $y^{(eq)}\mapsto y^{(eq)}$ is the fundamental thermodynamic relation in $(\NNN^{(eq)},0)$
\begin{eqnarray}\label{eq2.1}
S&=&S(E,N)\nonumber \\
E&=&E\nonumber \\
N&=&N
\end{eqnarray}
and the requirement (MaxEnt principle) that $S(E,N)$ reaches its maximum subjected to the constraint $E=E$ and $N=N$. We show that in this case the transformation $y^{(eq)}\mapsto y^{(eq)}$ is not a reduction but a one-to-one transformation. Indeed, let  $E^*$ and $N^*$ be Lagrange multipliers.  The maximization of $S(E,N)$ subjected to constraints $E=E$ and $N=N$ is made as follows: First, we introduce  a new potential (called a thermodynamic potential)
\begin{equation}\label{eq2.2}
\Phi(E,N;E^*,N^*)=-S(E,N)+E^*E+N^*N
\end{equation}
Second, we solve  $\Phi_E=0; \,\Phi_N=0$ (hereafter we use the notation $\Phi_x=\frac{\partial\Phi}{\partial x}$); let their solution be $\widehat{(E,N)}(E^*,N^*)$. Third, we introduce $S^*(E^*,N^*)=S(\widehat{(E,N)}(E^*,N^*);E^*,N^*)$, called a Legendre transformation of $S(E,N)$; $(E^*,N^*)$  are called conjugates of $(E,N)$.  By using the terminology and the notation that is standard in the equilibrium thermodynamics, $E^*=\frac{1}{T}$ and $N^*=-\frac{\mu}{T}$, where $T$ is the temperature and $\mu$ chemical potential. Finally, by making a  Legendre transformation of $S^*(E^*,N^*)$ we arrive at the initial entropy $S(E,N)$.

\subsubsection{$x\mapsto y^{(eq)}$}\label{eq20}

No restrictions are placed in this example on the model $(\MMM,\JJJ)$.
The point of departure of the transformation $\MMM\ni x\mapsto y^{(eq)}$ is the upper fundamental thermodynamic relation
\begin{eqnarray}\label{eq211}
S&=&S^{\uparrow}(x)\nonumber \\
E&=& E^{\uparrow}(x)\nonumber \\
N&=& N^{\uparrow}(x)
\end{eqnarray}
and the requirement (MaxEnt principle) that $S^{\uparrow}(x)$ reaches its maximum subjected to the constraint $E=E^{\uparrow}(x)$ and $N=N^{\uparrow}(x)$. The function $S^{\uparrow}:\MMM\rightarrow \mathbb{R}$ is called an upper entropy. The adjective ``upper'' indicates that its an entropy on the more microscopic (i.e. upper) level. The function $S^{\uparrow}(x)$ is assumed to be sufficiently regular and concave function. Similarly, $E^{\uparrow}(x)$ and $N^{\uparrow}(x)$, called an upper energy and an upper number of moles, are assumed to be sufficiently regular functions of $x$.

As in the previous example we introduce the upper thermodynamic potential
\begin{equation}\label{eq2.21}
\Phi^{\uparrow}(x;E^*,N^*)=-S^{\uparrow}(x)+E^*E^{\uparrow}(x)+N^*N^{\uparrow}(x)
\end{equation}
Let solutions to $\Phi^{\uparrow}_x=0$ be $\hat{x}(E^*,N^*)$. The quantity \\$S^*(E^*,N^*)=\Phi^{\uparrow}(\hat{x}(E^*,N^*);E^*,N^*)$ is the Legendre transformation of the thermodynamic relation $S^{\downarrow}(E,N)$ on the level of the equilibrium thermodynamics that is reduced from the thermodynamic relation (\ref{eq211}) on the level $(\MMM,\VVV)$. The transformation leading from the fundamental thermodynamic relation (\ref{eq211}) to the fundamental thermodynamic relation $S=S^{\downarrow}(E,N)$ is called a reducing Legendre transformation.

We illustrate the passage $x\mapsto y^{(eq)}$ on two well known examples. The first one is also historically the first.
The  fundamental thermodynamic relation $S=S^{\downarrow}(E,N)$ representing the ideal gas (obtained inside the classical equilibrium thermodynamics by making experiments, namely, by measuring the specific heat and the temperature-pressure-volume relation of dilute gases) has been derived by Boltzmann \cite{boltzmann}  from the kinetic theory (i.e. the theory in which the one particle distribution function $f(\rr,\vv)$ serves as the state variable; $\rr$ is the position vector and $\vv$ the momentum of one particle).   The Boltzmann fundamental thermodynamic relation (\ref{eq211}) on the level of kinetic theory is: $S^{\uparrow}(f(\rr,\vv))=-k_B\int d\rr\int d\vv f(\rr,\vv)\ln f(\rr,\vv)$; $k_B$ is the Boltzmann constant; $E^{\uparrow}(f(\rr,\vv)=\int d\rr \int d\vv \frac{\vv^2}{2m}$; $m$ is the mass of one particle; and $N^{\uparrow}(f(\rr,\vv)=\int d\rr\int d\vv f(\rr,\vv)$ (see e.g. \cite{pkg} for details of the calculations involved in the reducing Legendre transformation - see also Section \ref{3.1}).

The second example is the Gibbs equilibrium thermodynamics \cite{gibbscw}. The fundamental thermodynamic relation $S=S^{\downarrow}(E,N)$ of a given macroscopic system is obtained from the completely microscopic theory, i.e. a theory in which $n$-particle ($n\sim 10^{23}$) distribution function $f_n(\rr_1,\vv_1,...,\rr_n,\vv_n)$ serves as the state variable. In this microscopic theory the fundamental thermodynamic relation of the given macroscopic system is:\\ $S^{\uparrow}(f_n)=-k_B\int d\rr_1\int d\vv_1...\int d\rr_n\int d\vv_n f_n\ln f_n; \\E^{\uparrow}(f_n)=\int d\rr_1\int d\vv_1...\int d\rr_n\int d\vv_n e_n(\rr_1,\vv_1,...,\rr_n,\vv_n) f_n$, where  \\$e_n(\rr_1,\vv_1,...,\rr_n,\vv_n)$ is the microscopic energy (the microscopic Hamiltonian) of the given macroacopic system, and $N^{\uparrow}(f_n)= \int d\rr_1\int d\vv_1...\int d\rr_n\int d\vv_n f_n$ (see again \cite{pkg} for details of the calculations).

\subsection{$J\mapsto K$}\label{2.2}

We now proceed to static reductions in which both the initial and the reduced models involve  time evolution. There are essentially two avenues to follow. One, taken in \cite{dynmaxent}, follows closely Section \ref{eq20}. The upper fundamental thermodynamic relation (\ref{eq211}) is replaced by $S=S^{\uparrow}(x), y=Y^{\uparrow}(x)$. For example, if the upper level is kinetic theory and the lower level is hydrodynamics then $Y^{\uparrow}(x)$ are hydrodynamic fields expressed as first five moments (in $\vv$) of the one particle distribution function. The next step is to provide the submanifold that arises in the
MaxEnt transformation  with a vector field which is then the  reduced vector field $K$.  Several ways leading to $K$ are explored in \cite{dynmaxent}. In this paper we follow the second avenue on which the vector field $K$ itself arises in the MaxEnt reducing Legendre transformation.
We put our focus on the vector fields rather than on the state spaces as we did in  Section \ref{2.1} and in \cite{dynmaxent}. Formally, the static reduction is again a reducing Legendre transformation. The physical interpretations of the quantities entering it are however different. Moreover, the reduction $(\MMM,\JJJ)\rightarrow (\NNN,\KKK)$ can also be  made  in externally forced (or, in other words,  open) systems that cannot be reduced to $(\NNN^{(eq)},0)$ since the external forces prevent the approach to the thermodynamic equilibrium states. For example, in our recent work \cite{grmela2019entropy} we investigated the role of external forces in the case of heat conduction and how entropy and entropy production as potentials determining the evolution are related (just to highlight the difference note that vanishing total entropy production as a characterization of equilibrium state is insufficient).  The reduction $(\MMM,\JJJ)\rightarrow (\NNN,\KKK)$ brings to $(\NNN,\KKK)$ thermodynamics (we call it flux-thermodynamics) on the level $(\NNN,\KKK)$  even if on this level there is no  thermodynamics (i.e. there is no lower% \todo{neni tu spise lower?} 
fundamental thermodynamic relation since the passage $(\NNN,\KKK)\rightarrow (\NNN^{(eq)},0)$ cannot be made).
If however an upper fundamental thermodynamic relation on the level $(\NNN,\KKK)$ does exist then, as we shall see below, the quantities entering the reducing Legendre transformation   $(\MMM,\JJJ)\rightarrow (\NNN,\KKK)$ are closely related to rates of the quantities entering the reducing Legendre transformation $(\NNN,\KKK)\rightarrow (\NNN^{(eq)},0)$.

The point of departure for the investigation of the static reduction  $(\MMM,\JJJ)\rightarrow (\NNN,\KKK)$ is
the upper  fundamental flux-thermodynamic relation on the level  $(\MMM,\JJJ)$:
\begin{eqnarray}\label{eq0}
\mathfrak{S}&=&\mathfrak{S}^{\uparrow}(J)\nonumber\\
K&=& K^{\uparrow}(J)
\end{eqnarray}
We shall call $\mathfrak{S}^{\uparrow}$ an upper flux-entropy. We assume that it is a sufficiently regular and concave function of the fluxes $J$.

The reduction is made by the reducing Legendre transformation. This means that  we introduce  first the upper  flux-thermodynamic potential
\begin{equation}\label{eq01}
\Psi^{\uparrow}(J,K^{\dag})=-\mathfrak{S}^{\uparrow}(J)+\langle K^{\dag},K^{\uparrow}(J)\rangle
\end{equation}
where $K^{\dag}$ is the Lagrange multiplier (playing the role of the Lagrange multipliers $(E^*,N^*)$ introduced in the reducing Legendre transformations in the previous sections).

Let solutions to $\Psi^{\uparrow}_{J}=0$  be $\hat{J}(K^{\dag})$.
The quantity $\mathfrak{S}^{\downarrow \dag}(K^{\dag})=\Psi^{\uparrow}(\hat{J}(K^{\dag}),K^\dag)$ is then the  lower fundamental flux-thermodynamic  relation that is reduced from the  upper fundamental thermodynamic relation (\ref{eq0}). We note that $K=\mathfrak{S}^{\downarrow \dag}_{K^{\dag}}(K^{\dag})$ . The  reduced flux  $K$ and the flux $K^{\dag}$ introduced in the upper fundamental flux-thermodynamic relation (\ref{eq0})  are thus conjugate one to the other with respect to the lower flux-entropy $\mathfrak{S}^{\downarrow \dag}(K^{\dag})$. We recall that in the terminology  of the classical nonequilibrium thermodynamics the conjugates of the thermodynamic fluxes are called thermodynamic forces. Using this terminology, $K^{\dag}$ is the thermodynamic force corresponding to the thermodynamic flux $K$.

By making the reducing Legendre transformation  the ``unclosed'' flux $K^{\uparrow}(J)$ (unclosed since it depends on the upper vector field $J$) becomes ``closed'' via $\hat{J}(K^\dag)$, since it depends now on the Lagrange multiplier $K^{\dag}$ that we can freely choose. %\todo{vk: nerozumim: its value/relation is determined during the reduction, what do we mean by free choice here?}. 
If we choose it  to depend only on quantities belonging to the level $(\NNN,\KKK)$ then the flux $\KKK$ arising in the reducing Legendre transformation depends only on the quantities belonging to the level $(\NNN,\KKK)$. The problem of the ``closure'' of  $K(J)$ was thus transformed  into the problem of the specification of the Lagrange multiplier $K^{\dag}$. We shall see in Section \ref{3}  this reformulation of the problem of the closure in the context of the pattern-recognition type analysis of the time evolution in $(\MMM,\JJJ)$. Here we limit ourselves only  to a formal specification of $K^{\dag}$. In the case of externally forced systems, the thermodynamic forces $K^{\dag}$ are often the external forces. In the case of externally unforced systems (i.e. the systems that can reach the level of the equilibrium thermodynamics) the upper entropy $S^{\uparrow})$ (appearing in the fundamental thermodynamic relation (\ref{eq211}) with $x$ replaced by $y$) exists,  we shall choose $K^{\dag}$ in such a way that: (i) $K^{\dag}$ is a function of $y^*=S^{\uparrow}_{y}(y)$, and (ii) $\langle y^*,\mathfrak{S}^{\downarrow \dag}_{y^*}\rangle=a \langle K^{\dag},\mathfrak{S}^{\downarrow \dag}_{K^{\dag}}\rangle$, where $a$ is a real positive number.
This then means that (in the case of externally unforced systems for which the upper entropy $S^{\uparrow}$ exists)%\todo{VK: omlouvam se, ted v rychlosti to tu neumim opravit, nevychazi mi stale znamenko. Totiz myslim, ze: jeli $\sigma^\uparrow$ konkavni, tak by vychazela $\dot{S}<0$ v (7) (bylo minus v rovnici pro $\dot{\rho}$ v posledni vete teto sekce, nevim proc, bylo obracene znamenko v difuzni rovnici a opet $\dot{S}^\uparrow<0$). Kdyz ale vezmu sigmu konvexni, tak musim vzit opacne znamenko v $\mathfrak{S}^{\uparrow}(J)$ a nasledne vsechno sedi az na tu difuzni rovnici - vyjde s minusem, coz je blbost.  Neco tu delam blbe..}
\begin{equation}\label{eq23}
\dot{S^{\uparrow}}=-\langle y^*,\mathfrak{S}^{\downarrow \dag}_{y^*})\rangle = -a\langle K^{\dag},\mathfrak{S}^{\downarrow \dag}_{K^{\dag}}\rangle >0
\end{equation}
provided
\begin{equation}\label{eq02}
\dot{y}=-\mathfrak{S}^{\downarrow \dag}_{y^*}=-K(y)
\end{equation}
Here we see the physical interpretation of $\mathfrak{S}^{\downarrow }(y)$. The rate of the upper entropy  $S^{\uparrow}(y)$ equals $a\langle K^{\dag},\mathfrak{S}^{\downarrow \dag}_{K^{\dag}}\rangle$. We emphasize again that this interpretation applies only in the case when the entropy $S^{\uparrow}(y)$ exists, i.e. in the case of externally unforced systems.

Both in the case of externally forced and unforced systems, the flux-entropy $\mathfrak{S}^{\downarrow }(y)$ is the new addition to the model $(\NNN,\KKK)$ arriving from putting it into the context of  a more microscopic model $(\MMM,\JJJ)$. The reduction made above is only formal, its physical basis will be discussed in Section \ref{3}.

We end this section with a simple illustration. We choose the upper level $(\MMM,\JJJ)$ with $\MMM\ni x=f(\rr,\vv)$  and the flux $J(\rr,\vv)=(J_1(\rr,\vv), J_2(\rr,\vv),J_3(\rr,\vv))$. The upper fundamental flux-thermodynamic relation   is $\mathfrak{S}^{\uparrow}(J)= -\frac{1}{2}\Lambda \int d\rr\int d\vv f J_iJ_i$,  $K_i=\int d\vv f J_i$,  and  $K^{\dag}_i(\rr)=\frac{\partial\rho^*}{\partial r_i}$, where $\rho(\rr)=\int d\vv f(\rr,\vv)$ is the state variable on the lower level $(\NNN,\KKK)$,  $\rho^*(\rr)=S^{\uparrow}_{\rho(\rr)}(\rho)$, and $\Lambda>0$ is a parameter. We use the notation: $i=1,2,3$ and  the summation convention over the repeated indices. Simple calculations show that $J_j=-\frac{1}{\Lambda} \frac{\partial \rho^*}{\partial r_i}$, $\mathfrak{S}^{\downarrow \dag}(K^{\dag})=-\frac{1}{2 \Lambda}\int d\rr \rho K^{\dag}_i K^{\dag}_i$,
%\todo{proc je tu minus ve vztahu pro derivaci rho? Myslim, ze tu byt nema.}
$\dot{\rho}=-\mathfrak{S}^{\downarrow \dag}(K^{\dag}(\rho^*))_{\rho^*}=-\frac{\partial}{\partial r_i}\left(\frac{\rho}{\Lambda}\frac{\partial \rho^*}{\partial r_i}\right)$ is the diffusion equation, %\todo{tady v difuzni rovnici by minus byt nemelo}
and $\dot{S}^{\uparrow}=\int d\rr\frac{\rho}{\Lambda}\frac{\partial \rho^*}{\partial r_i}\frac{\partial \rho^*}{\partial r_i}>0$ is the entropy production.% \todo{vychazi mi zaporne}.%jen pro VK: rho^*=-chem pot, takze ok

\section{Dynamic Reductions}\label{3}

We turn  now to the questions of where do the fundamental thermodynamic relations come from and why it is the reducing Legendre transformation  that makes   the reduction. Answers to both questions must come from a detailed investigation of solutions to the governing equations on the level $(\MMM,\JJJ)$ (i.e. a detailed investigation of trajectories generated by $(\MMM,\JJJ)$). Such investigation consists of three steps: (\textit{Step 1}) generating $\mathcal{P}$, i.e. solving the governing equations of
the model $(\MMM,\JJJ)$; the phase portrait $\mathcal{P}$ serves then as
the data base for the further investigation in the next two steps. (\textit{Step 2}) recognizing a pattern $\mathcal{R}$ in $\mathcal{P}$.  (\textit{Step 3}) Identifying a model $(\NNN,\KKK)$ for which $\mathcal{R}$ recognized in  Step 2 is its phase portrait. All three steps are obviously very difficult to make. Following the experience collected in investigations of particular examples of reductions (in particular the BBGKY and Grad hierarchies \cite{marsden-bbgky}, \cite{grad,miroslav-grad}, \cite{rugjap}, the Chapman-Enskog method \cite{dgm} or thermodynamics with internal variables \cite{van-berezovski}),  we suggest below  a general strategy for the dynamic reduction.

First, in Section \ref{3.1}, we show that the time evolution that makes most directly the reducing Legendre transformations is the gradient dynamics. In Section \ref{3.2} we begin with a less formal (more physically justified) dynamics, namely with the Hamiltonian dynamics. In order to prepare it for the pattern recognition process $\mathcal{P}\rightarrow \mathcal{R}$ we reformulate it first into a hierarchy (that we call a Poisson-Grad hierarchy) that preserves the Hamiltonian kinematics. Subsequently, by adding an appropriate dissipation term,   the Hamiltonian vector field is transformed  into a GENERIC vector field. Finally,
the  viewpoint developed originally in the Chapman-Enskog analysis is  used to solve approximately the GENERIC Poisson hierarchy and arrive at the reducing Legendre transformation.

\subsection{Gradient dynamics}\label{3.1}

What are the vector fields $\JJJ$ that are compatible with the reduction $x\rightarrow y^{(eq)}$ made in Section (\ref{eq20})? In other words, what is the time evolution that, by following it to its conclusion, makes the transformation $x\rightarrow \hat{x}(E^*,N^*)$ introduced  in Section (\ref{eq20})? One obvious candidate \cite{ch}, \cite{landau-ginzburg} is the gradient time evolution governed by
\begin{equation}\label{eqgr}
\dot{x}=-\Lambda \Phi^{\uparrow}_x
\end{equation}
where $\Lambda$ is a positive definite operator. Indeed, (\ref{eqgr}) implies $\dot{\Phi}^{\uparrow}=-\langle \Phi^{\uparrow}_x,\Lambda\Phi^{\uparrow}_x\rangle <0$.
This means that the thermodynamic potential $\Phi^{\uparrow}$ plays the role of the Lyapunov function %\todo{vk: muzu se pokusit sepsat tu odstavec o korektnejsim pohledu na Lyapunova v takto obecnych rovnicich pomoci semigrup; ale je to v podstate nanic, jen by to asi lepe vypadalo (pro matematika).}
for the approach $x\rightarrow \hat{x}(E^*,N^*)$ (we recall that we have assumed already in the previous sections that $\Phi^{\uparrow}$ is a convex function of $x$).  This means that by following the time evolution governed by (\ref{eqgr}) to its conclusion (i.e. $t\rightarrow \infty$) we are making the reducing Legendre transformation $S^{\uparrow}(x)\rightarrow S^*(E^*,N^*)$.
If the operator $\Lambda$ is degenerate in the sense that $\Lambda E^{\uparrow}_x=0$ and $\Lambda N^{\uparrow}_x=0$ then the time evolution governed by (\ref{eqgr}) can be seen as the maximization of the entropy $S^{\uparrow}(x)$ subjected to constraints $E=E^{\uparrow}(x)$ and $N=N^{\uparrow}(x)$. Historically, the role of the gradient dynamics in reductions to the equilibrium has been recognized in \cite{ch}, \cite{landau-ginzburg}.

We  note that (\ref{eqgr}) can be replaced by a more general gradient time evolution governed by
\begin{equation}\label{eqgrr}
\dot{x}=-[\Xi^{\uparrow}_{x^*}(x,x^*)]_{x^*=\Phi^{\uparrow}_x}
\end{equation}
provided $\Xi^{\uparrow}(x,x^*)$, called an upper dissipation potential,
is a sufficiently regular real valued function satisfying the following three properties: (i) $\Xi^{\uparrow}(x,0)=0$, (ii) $\Xi^{\uparrow}_{x^*}(x,x^*)]_{x^*=0}=0$, and (iii) $\Xi^{\uparrow}_{x^*}(x,x^*)$ is a convex function $x^*$ in a neighborhood of $x^*=0$. Indeed, in the time evolution governed by (\ref{eqgrr}) the thermodynamic potential $\Phi^{\uparrow}$  plays also the role of the Lyapunov function since  $\dot{\Phi}^{\uparrow}=-[\langle x^*,\Xi^{\uparrow}_{x^*}\rangle]_{x^*=\Phi^{\uparrow}_x} <0$ due to the  three properties that the upper dissipation potential $\Xi^{\uparrow}$ is required to satisfy.
If in particular $\Xi^{\uparrow}(x,x^*)=\frac{1}{2}\langle x^*,\Lambda x^*\rangle$ then (\ref{eqgrr}) turns into (\ref{eqgr}). The vector field $\JJJ$ in (\ref{eqgrr}) (i.e. the right hand side of (\ref{eqgrr})) is thus a direct generalization of the vector field $\JJJ$ in (\ref{eqgr}). With an additional requirement that $\Xi^{\uparrow}$ is degenerate in the sense that the energy $E(f)$ and the number of moles $N(f)$ are its dissipative Casimirs \footnote{We say that $C^{(diss)}(x)$ is a dissipative Casimir of the dissipative potential $\Xi^{\uparrow}(x,x^*)$ if
$\langle C^{(diss)}_x,\Xi^{\uparrow}_{x^*}\rangle=0$ and $\langle x^*,[\Xi^{\uparrow}_{x^*}]_{x^*=C^{(diss)}_x}\rangle=0$.} then the time evolution governed by (\ref{eqgrr}) maximizes the entropy $S^{\uparrow}(x)$ subjected to constraints $E=E^{\uparrow}(x)$ and $N=N^{\uparrow}(x)$.

Summing up, with the gradient dynamics we are making only a small step towards understanding the physical basis of the static reduction. We learned  that the upper entropy plays the role of the potential generating the approach to the reduced pattern $\mathcal{R}$. The upper entropy  is therefore a quantity that comes from the information collected about the way the pattern $\mathcal{R}$ is emerging in the phase portrait $\mathcal{P}$. The reducing Legendre transformation is then a mathematical formulation of the fact that the upper thermodynamic potential $\Phi^{\uparrow}$ plays the role of the Lyapunov function in the emergence of the pattern $\mathcal{R}$.
 We still do not know, however, why is the reducing time evolution  governed by the gradient dynamics.

\subsection{Hamiltonian dynamics}\label{3.2}

In order to enter deeper into the physics of the reduction, we
have to turn to mechanics. This is because the time evolution that takes place on the most microscopic level $(\MMM^{(micro)},\JJJ^{(micro)})$ is governed by the classical mechanics, The  mechanics  is then expected to provide the physical basis also for    more macroscopic dynamical theories. In this paper   we do not consider  more microscopic theories in which  quantum mechanics has to  replace the   classical mechanics. From the mathematical point of view,
we choose  the Hamiltonian formulation  of the classical mechanics. We are making this choice  because the Hamiltonian   formulation  has proven to be particularly useful  in attempts to combine mechanics   with other theories (e.g. with geometric optics or with thermodynamics), in attempts to extend mechanics (e.g. to  quantum mechanics), and in attempts to recognize the geometry involved in mechanics (e.g the symplectic or the contact geometries).
The continuum version of mechanics (represented in  the Euler equation) has been put into the Hamiltonian form by Clebsch \cite{clebsch} and later, by Arnold \cite{arnold} where the connection of non-canonical Hamiltonian structures with the Lie group theory was recognized. As in the particle mechanics, the usefulness of the Hamiltonian formulation of the continuum mechanics has come into light in particular in extensions, in unifications with other mesoscopic theories, in relations to thermodynamics, in numerical solutions, and in geometrical formulations (see \cite{pkg} and references cited therein).

Advantages of the Hamiltonian formulation stem mainly from the fact that the vector field generating the time evolution involves two objects that have two different and  independent physical contents. One is the geometrical structure expressing mathematically kinematics of the chosen state variables and the other is a potential (a real valued function) representing the energy (i.e. the quantity involving all the internal mechanical forces). In reductions we consider the kinematics and the energy separately. This is the main contribution (and advantage) of the dynamic reductions discussed below.

The Hamiltonian time evolution of $x\in \MMM$ is governed by
\begin{equation}\label{Ham1}
\frac{\partial x}{\partial t}=LE_x
\end{equation}
$E(x)$ is the energy and $L$ is a Poisson bivector expressing mathematically the kinematics of $x$.
The vector field appearing on the right hand side of  (\ref{Ham1}) is thus a covector $E_x$ transformed into a vector by the kinematics which is mathematically expressed in the Poisson bivector $L$.

A bivector $L$ is a Poisson bivector if  the bracket
\begin{equation}\label{br1}
\{A,B\}=\langle A_x,LB_x\rangle
\end{equation}
is a Poisson bracket. By $\langle.\rangle$ we denote  the pairing in the space with $x$ as its  elements, $A$ and $B$ are real valued and sufficiently regular functions of $x$. A bracket $\{A,B\}$ is a Poisson bracket if the following relations hold: $\{A,B\}=-\{B,A\}$, and $\{\{A,B\},C\}+\{\{B,C\},A\}+\{\{C,A\},B\}=0$.
We note that with the bracket (\ref{br1}) the time evolution equation (\ref{Ham1}) can  alternatively be written in the form
\begin{equation}\label{Ham2}
\frac{dA}{dt}=\{A,E\}; \forall A
\end{equation}
An important property of $L$ is its degeneracy. We call a non constant real valued function $C(x)$ a Casimir if
\begin{equation}\label{Casimir}
\{A,C\}=0; \forall A
\end{equation}
We shall see later   that,  from the physical point of view, the Casimir functions have the interpretation of various types of  entropies.

From the properties of $L$ listed above, we can immediately deduce the following properties of  solutions to (\ref{Ham1}):
\begin{eqnarray}\label{En}
\frac{dE}{dt}&=&0\\ \label{Ent}
\frac{dC}{dt}&=&0
\end{eqnarray}
Equation (\ref{En}) expresses the energy conservation. It is a direct consequence of (\ref{Ham2}) and the property $\{A,B\}=-\{B,A\}$. Indeed, $\dot{E}=\{E,E\}=0$. Equation (\ref{Ent}) expresses the entropy conservation and is a direct consequence of (\ref{Ham2}) and the degeneracy (\ref{Casimir}). We note that both the energy $E$ and the Casimirs $C$ are conserved but for two very different reasons. The former because $E$ is the generating potential and the Poisson bracket is skewsymmetric, the latter because of the degeneracy of the Poisson bracket (i.e. the degeneracy of the kinematics).

In our attempt to contribute to the clarification of the physics involved in the pattern-recognition type passage $(\MMM,\JJJ) \rightarrow (\NNN,\KKK)$, we take $(\MMM,\JJJ)$ to be the kinetic theory. We therefore present now the Hamiltonian structure of this theory.

The kinematics of the one particle distribution function $f(\rr,\vv)$ that serves as the state variable in kinetic theory is induced from the kinematics of one particle in the classical mechanics, i.e. from the Lie group of transformations $(\rr,\vv)\mapsto (\rr',\vv')$ preserving the Poisson bracket $\{a,b\}=\frac{\partial a}{\partial r_i}\frac{\partial b}{\partial v_i}- \frac{\partial b}{\partial r_i}\frac{\partial a}{\partial v_i} $;  $a$ and $b$ are real valued functions of $(\rr,\vv)$. Such transformations are called in classical mechanics canonical transformation. The  path: \textit{(Lie group)} $\rightarrow$ \textit{(the corresponding to it  Lie algebra)} $\rightarrow$ \textit{(its dual)} $\rightarrow$ \textit{(Poisson bracket on the dual of the Lie algebra induced by the structure of the Lie group)},  that is an integral part of the theory of Lie groups \cite{arnold},  leads to the Poisson bracket
\begin{equation}\label{br2}
\{A,B\}^{(k)}=\int d\rr\int d\vv f\left(\frac{\partial A_f}{\partial r_i}\frac{\partial B_f}{\partial v_i}-\frac{\partial B_f}{\partial r_i}\frac{\partial A_f}{\partial v_i}\right)
\end{equation}
expressing mathematically the kinematics of the one particle distribution function $f(\rr,\vv)$. We use hereafter the summation convention. Regarding the  degeneracy of (\ref{br2}),
\begin{equation}\label{Casimir2}
S(f)=\int d\rr\int d\vv \eta(f)
\end{equation}
where $\eta(f)$ is a sufficiently regular function $\eta:\mathbb{R}\rightarrow\mathbb{R}$, are all Casimirs of (\ref{br2}). A simple direct verification of (\ref{Casimir}) proves it.

With the Poisson bracket (\ref{br2}), the kinetic equation (\ref{Ham1}) becomes
\begin{equation}\label{kin1}
\frac{\partial f}{\partial t}=-\frac{\partial}{\partial r_i}\left(f\frac{\partial E_f}{\partial v_i}\right)+\frac{\partial}{\partial v_i}\left(f\frac{\partial E_f}{\partial r_i}\right)
\end{equation}
This equation (in fact a family of equations parametrized by the energy $E(f)$) is the point of departure. First, in Section \ref{3.3} we recall the Boltzmann analysis of the approach to equilibrium and in Section \ref{3.4} we discuss the approach to fluid mechanics.

\subsection{Poisson hierarchies}\label{PH}

Before discussing reductions in Hamiltonian systems, we  turn to a less ambitious goal. We just want to reformulate  the Hamiltonian dynamics into a new form  that may  hopefully be more suitable for the pattern recognition process in the phase portrait $\mathcal{P}$. As for the passage $f(\rr,\vv)\rightarrow y^{(eq)}$, we shall see that a useful reformulation (due to Boltzmann \cite{boltzmann}) consists of  identifying  one particular event in the time evolution, namely the binary collision, and separating  the Hamiltonian vector field into two parts, one generating the outcome of binary collisions and the other the rest of the time evolution. The Hamiltonian vector field generating the binary collisions is then modified into a gradient vector field discussed in Section \ref{3.1}. The physical justification of the modification is the ignorance of  details of the complex trajectories of colliding particles.

This Boltzmann's insight is not however adequate to investigate the  reduction $J\rightarrow K$ leading from the kinetic theory to  fluid mechanics (that  is a level on which a reduced time evolution takes place,  a level  that is less detailed than the level of kinetic theory but more detailed than the level of the equilibrium thermodynamics). We shall  use Grad's insight \cite{grad} to make  a reformulation suitable for this type of investigation.   We however use   Grad's insight to reformulate only the   kinematics (the Poisson bracket (\ref{br2})). The resulting reformulation, that we call Poisson-Grad hierarchy, is thus different from the Grad reformulation known as Grad hierarchy. The Poisson-Grad hierarchy
provides a Hamiltonian kinetic equation, that, if modified in a similar way as Boltzmann has modified (\ref{kin1}), becomes a kinetic equation providing dynamical basis for  the static reduction  $\JJJ\rightarrow \KKK$ (see Section \ref{2.2}), where $\JJJ$ are the vector fields of kinetic theory and  $\KKK$  the vector fields of fluid mechanics.

Following Grad, we begin the reduction by anticipating that the state variables of fluid mechanics
$(\rho(\rr),\uu(\rr),s(\rr))$ are expressed in terms of $f(\rr,\vv)$ as follows:
\begin{eqnarray}\label{Mfl}
\rho(f;\rr)&=&\int d\vv f(\rr,\vv)\nonumber \\
\uu(f;\rr)&=&\int d\vv \vv f(\rr,\vv)\nonumber \\
s(f;\rr)&=&\int d\vv \eta(f(\rr,\vv))\nonumber \\
\end{eqnarray}
The field  $\rho$ is the field of the mass density, $\uu$ the momentum  density, and $s$ the entropy density. Instead of the entropy field $s(\rr)$ we could also choose the energy field $e(\rr)$. We shall discuss the difference later.
We could include also other fields as e.g. the entropy flux, the stress tensor etc. With such extended  set of state variables, the fluid mechanic becomes an extended fluid mechanics.  All the steps that we shall make below with the fields  $(\rho(\rr),\uu(\rr),s(\rr))$ would remain  unchanged,   only the calculations and the resulting equations would be more complex.
The quantity $\eta(f(\rr,\vv))$ is the quantity introduced in (\ref{Casimir2}). At this point we leave it unspecified.

 The  relation   (\ref{Mfl}) between $(\rho(\rr),\uu(\rr),s(\rr))$ and $f(\rr,\vv)$ is  based on the physical interpretation of these state variables. Such (or similar) relations   should however  arise in the process of recognizing the pattern $\mathcal{R}$ (representing the phase portrait of fluid mechanics) in the phase portrait $\mathcal{P}$ of kinetic theory. They
should not be imposed at the beginning of the pattern recognition process. In this paper we however begin the pattern recognition process with (\ref{Mfl}).

The next step is the key step in the reformulation. Our objective is to reformulate the kinetic theory kinematics, i.e. the Poisson bracket (\ref{br2}). In  (\ref{br2}), we consider $A(f)$ and $B(f)$ to depend on $f$ in two ways.  First, $A(f),B(f)$ depend on $f$  in the same way as in (\ref{br2}) and second, through their  dependence on $(\rho(\rr),\uu(\rr),s(\rr))$ that are related to $f$ in (\ref{Mfl}). The state variables are now
\begin{equation}\label{svPG}
x=(\rho(\rr),\uu(\rr),s(\rr), f(\rr,\vv))
\end{equation}
From the physical point of view, we regard now the fields $(\rho(\rr),\uu(\rr),s(\rr))$ as the principal state variables and $f(\rr,\vv)$ as a variable expressing extra details. We can interpret $f(\rr,\vv)$ as ``fluctuations'' but we  do not use in this paper the  tools of stochastic formulations.

 We arrive at the Poisson bracket expressing kinematics of (\ref{svPG})   by replacing $A_f$ appearing in the Poisson bracket (\ref{br2})  with $A_f+A_{\rho}+v_iA_{u_i}+\eta_fA_s$ and $B_f$ with $B_f+B_{\rho}+v_iB_{u_i}+\eta_fB_s$. In other words, we extend the class of functions $A$ and $B$ in (\ref{br2}) to those that depend on $f$ also through their dependence on $(\rho(f),\uu(f),s(f))$ given in (\ref{Mfl}).
After straightforward calculations we arrive at
\begin{equation}\label{brfl}
\{A,B\}^{(PG)}=\{A,B\}^{(kt)}+\{A,B\}^{(fl)}+\{A,B\}^{(ktfl)}
\end{equation}
where $\{A,B\}^{(kt)}$ is the kinetic theory Poisson bracket (\ref{br2}),
\begin{eqnarray}\label{flbr}
\{A,B\}^{(fl)}&=&\int d\rr\int d\vv \left[\rho\left(\frac{\partial A_{\rho}}{\partial r_i}B_{u_i}-\frac{\partial B_{\rho}}{\partial r_i}A_{u_i}\right)\right.\nonumber \\
&&\left.+s\left(\frac{\partial A_{s}}{\partial r_i}B_{u_i}-\frac{\partial B_{s}}{\partial r_i}A_{u_i}\right)\right.\nonumber \\
&&\left.+u_i\left(\frac{\partial A_{u_i}}{\partial r_j}B_{u_j}-\frac{\partial B_{u_i}}{\partial r_j}A_{u_j}\right)\right]
\end{eqnarray}
and %\todo{(ktfl) vs (ffl) - vzal jsem tu prvni}
\begin{eqnarray}\label{ktfl}
\{A,B\}^{(ktfl)}&=&\int d\rr\int d\vv \left[f\left(\frac{\partial A_f}{\partial r_i}B_{u_i}-\frac{\partial B_f}{\partial r_i}A_{u_i}\right)\right.\nonumber \\
&&\left.+f\frac{\partial \eta_f}{\partial v_i}\left(\frac{\partial A_f}{\partial r_i}B_{s}-\frac{\partial B_f}{\partial r_i}A_{s}\right)\right.\nonumber \\
&&\left.+f\left(\frac{\partial A_{\rho}}{\partial r_i}\frac{\partial B_f}{\partial v_i}-\frac{\partial B_{\rho}}{\partial r_i}\frac{\partial A_f}{\partial v_i}\right)\right.\nonumber \\
&&\left.+fv_j\left(\frac{\partial A_{u_j}}{\partial r_i}\frac{\partial B_f}{\partial v_i}-\frac{\partial B_{u_j}}{\partial r_i}\frac{\partial A_f}{\partial v_i}\right)\right.\nonumber \\
&&\left.+f\left(\frac{\partial (A_s\eta_f)}{\partial r_i}\frac{\partial B_f}{\partial v_i}-\frac{\partial (B_s\eta_f)}{\partial r_i}\frac{\partial A_f}{\partial v_i}\right)\right]
\end{eqnarray}

The time evolution equations (\ref{Ham1}) with the Poisson bracket (\ref{brfl}) and the energy (\ref{enfl})
are
\begin{eqnarray}\label{eqs11}
\frac{\partial}{\partial t}\left(\begin{array}{ccc}\rho\\u_i\\e\end{array}\right)&=&-\left(\begin{array}{ccc}\frac{\partial (\rho E_{u_i})}{\partial r_i}\\\frac{\partial (u_iE_{u_j})}{\partial r_j}+\frac{\partial p}{\partial r_i}\\ \frac{\partial [(\epsilon +p)E_{u_i}]}{\partial r_i}\end{array}\right)-\int d\vv \left(\begin{array}{ccc}\frac{\partial \left( f\frac{\partial E_f}{\partial v_i}\right)}{\partial r_i}\\\frac{\partial \left( fv_i\frac{\partial E_f}{\partial v_j}\right)}{\partial r_j}
\\ \frac{\partial (f E_f E_{u_i})}{\partial r_i}+\frac{\partial \left(\Pi\frac{\partial E_f}{\partial v_i}\right)}{\partial r_i}\end{array}\right)\nonumber\\
\frac{\partial f}{\partial t}&=& -\frac{\partial}{\partial r_i}\left[f\left(E_{u_i}+\frac{\partial\eta_f}{\partial v_i}E_s\right)\right] \nonumber \\
&&+\frac{\partial}{\partial v_i}\left[f\left(\frac{\partial E_{\rho}}{\partial r_i}+\frac{\partial (\eta_fE_s)}{\partial r_i}+v_j\frac{\partial E_{u_j}}{\partial r_i}\right)\right]\nonumber \\
&&-\frac{\partial}{\partial r_i}\left(f\frac{\partial E_f}{\partial v_i}\right)+\frac{\partial}{\partial v_i}\left(f\frac{\partial E_f}{\partial r_i}\right)\nonumber \\
\end{eqnarray}
where $p=-e +\rho E_{\rho}+u_i E_{u_i}+sE_s +\int d\vv fE_f=-\epsilon+\rho E_{\rho}+u_i E_{u_i}+sE_s$ is the scalar hydrodynamic pressure  and $\Pi=f E_{\rho}+\eta E_s+ fE_f$. In addition, the equation governing the time evolution of the entropy field $s(\rr)$ is
\begin{equation}\label{seq}
\frac{\partial s}{\partial t}=-\frac{\partial (sE_{u_i})}{\partial r_i}-\frac{\partial}{\partial r_i}\left(\int d\vv \eta\frac{\partial E_f}{\partial v_i}\right)
\end{equation}

The energy $E$ in (\ref{eqs11}) is, at this point, completely arbitrary
\begin{equation}\label{enfl}
E(f,\rho,\uu,s)=\int d\rr e(f,\rho,\uu,s,f;\rr)
\end{equation}
Also the function $\eta(f)$ appearing in (\ref{eqs11}) is an unspecified function $\eta:\mathbb{R}\rightarrow \mathbb{R}$ (see (\ref{Casimir2})).

We call the  time evolution equations (\ref{eqs11}) a Poisson-Grad hierarchy, ``Grad'' because they couple the  Euler fluid mechanics equations to a more microscopic description of fluids, and ``Poisson'' because  they retain the Poisson kinematics of both fluid mechanics and kinetic theory. In the Grad hierarchy \cite{grad} the Euler equations are coupled to the higher order moments (in the momentum  variable $\vv$) of $f(\rr,\vv)$, the energy $E$ is fixed, it is  the kinetic energy $\int d\rr\int d\vv \frac{\vv^2}{2m}$ (i.e. the fluid described by the Euler equations is an ideal gas), and the equation governing the time evolution of the entropy $s(\rr)$ is absent. On the other hand, the Poisson-Grad hierarchy involves an unspecified energy (\ref{enfl}). This means that the Poisson-Grad hierarchy  addresses  general fluids and not only ideal gases. The presence of the equation governing the time evolution of the entropy field (\ref{seq}) is another important contribution of the Poisson-Grad hierarchy.
The Euler part in the Poisson-Grad hierarchy (i.e. the first equation in (\ref{eqs11}) without the second term on its right hand side) is still coupled to $f$ since the energy $E$ depends on $f$. This coupling can be however easily removed by a special choice of the energy $E$. We note that if  the energy $E$ is a sum of two terms, one depending only on the hydrodynamic fields and the other depending only on $f$ then the Euler part becomes completely decoupled.

 The infinite version of the Poisson-Grad hierarchy, i.e. the version  in which $f(\rr,\vv)$ in (\ref{svPG}) is replaced by an infinite number of higher moments, has been worked out  in \cite{miroslav-grad}. We can indeed interpret $f(\rr,\vv)$ in (\ref{eqs11}) as representing infinite number of higher moments. Contrary to the Grad hierarchy, the second equation in (\ref{eqs11}) (i.e. the kinetic equation) includes explicitly the coupling to the hydrodynamic fields. The second term on the right hand side of the first equation in (\ref{eqs11}) can also be interpreted  as an analogue of the Langevin term expressing the influence of a ``noise'' on the fluid motion. But then the second equation in (\ref{eqs11}) is the equation describing the time evolution of such ``noise''.  In the standard stochastic formulation the noise is imposed and fixed.

\subsection{GENERIC dynamics: $x\rightarrow y^{(eq)}$ as $t\rightarrow \infty$}\label{3.3}

Now we begin  the reduction process in the Hamiltonian dynamics. In this section we recall   the Boltzmann  passage $x\rightarrow y^{(eq)}$ as $t\rightarrow \infty$. We cast it into the general viewpoint suggested in this paper.

Following our general strategy, the first step in  the passage $\mathcal{P}\rightarrow \mathcal{R}$  is to generate  the phase portrait $\mathcal{P}$ corresponding to the kinetic time evolution (\ref{kin1}). While it is  possible, at least in principle, to make direct simulations with  contemporary  computers and get some information about $\mathcal{P}$ in this way, we shall use for this purpose  Boltzmann's insight. If the macroscopic systems under investigation are rarefied gases then the gas-particle trajectories will have a complex texture due to their larger changes occurring in collisions.  The complexity  of $\mathcal{P}$ is expected to be essential for the emergence of the pattern $\mathcal{R}$ corresponding to the level of equilibrium thermodynamics  in which all details  are erased, only
the total energy $E$,  the total number of moles $N$,  and one other  feature (that is inherited from the pattern-emergence process and that finds its mathematical formulation in the fundamental thermodynamic relation $S=S(E,N)$) remain.  Following Boltzmann's insight, the main culprit of the complexity in the  texture of $\mathcal{P}$ are collisions.

In order to make the pattern emergence manifestly visible in solutions to (\ref{kin1}), Boltzmann has  modified the Hamiltonian kinetic  equation   by adding to its right hand side the term (\ref{eqgrr}) in which $x=f(\rr,\vv)$. The new added term represents  the contribution of collisions to the time evolution. Boltzmann's idea of making the equilibrium pattern $\mathcal{R}$ visible  is to replace $\frac{\partial f}{\partial t}$ with $\left(\frac{\partial f}{\partial t}\right)_{free\,flow}+\frac{\triangle f}{\triangle t}$, where the  first term is the vector field generating trajectories of non-colliding gas particles and the second term (collision term) is the contribution of collisions.
The particle trajectories entering and leaving the collisions are first seen in their completeness and then they are represented as generated by  the vector field $\frac{\triangle f}{\triangle t}$. The local details of the trajectories of colliding particles are ignored. In other words, the Hamiltonian vector field governing the time evolution of binary collisions is replaced by a new vector field $\frac{\triangle f}{\triangle t}$  that is obtained by, first, letting the original vector field to generate the trajectories, second, selecting only some important features of the trajectories, and third, constructing a new vector field generating the selected features of the trajectories.
Such procedure, used somewhat implicitly by Boltzmann, has been explicitly suggested in \cite{ehrenfests,gk-ehrenfest} and called in \cite{ehrenfest-regularization} Ehrenfest regularization.
If the Boltzmann collision term $ \frac{\triangle f}{\triangle t}$ (obtained by Boltzmann by analyzing the mechanics of binary collisions) is  cast into the form (\ref{eqgrr}) (we denote such dissipation potential by the symbol $\Xi^{(Boltzmann)}(f,f^*)$ - see details in \cite{pkg}),  the Boltzmann entropy (appearing in  Section \ref{eq20}) appears as a result.

 The time evolution governed by the Boltzmann kinetic equation is indeed  entailing the reducing Legendre transformation $(\MMM,\VVV)\rightarrow (\NNN^{(eq)},0)$ discussed in Section \ref{eq20}. The thermodynamic potential $\Phi(f;E^*,N^*)$ plays the role of the Lyapunov function for the approach $f(\rr,\vv)\rightarrow \hat{f}(\rr,\vv;E^*,N^*)$ as $t\rightarrow \infty$ since the Boltzmann entropy $S(f)$ and the number of moles $N(f)$ are  Casimirs of the Poisson bracket (\ref{br2}) and the energy $E(f)$ (only the kinetic energy in the case of the Boltzmann equation) and the number of moles $N(f)$ are  dissipative Casimirs of the Boltzmann dissipative potential $\Xi^{(Boltzmann)}$. The pattern $\mathcal{R}$ expressing the level of the equilibrium thermodynamics in the phase portrait $\mathcal{P}$ corresponding to the Boltzmann kinetic equation is composed of the
distribution functions $\hat{f}(\rr,\vv;E^*,N^*)$ called the total Maxwell distribution functions. With the Boltzmann analysis sketched above, we have seen where does the fundamental thermodynamic relation (\ref{eq211}) come from ($E^{\uparrow}(f)$ and $N^{\uparrow} (f)$ are constants of motion,  and $S^{\uparrow}(f)$ is the potential driving the approach to fixed points)  and also why MaxEnt appears (the upper thermodynamic potential $\Phi^{\uparrow}(f;E^*.N^*)$ plays the role of the Lyapunov function in the approach to fixed points).

The Hamiltonian formulation of kinetic equations presented in Section \ref{3.2} allows us to bring  Boltzmann's analysis into  a more abstract setting  and then use it in a larger context. In particular: (i) we have learned  that we have to  look for the entropy $S^{\uparrow}(x)$ in Casimirs of the Poisson bracket expressing the kinematics of the Hamiltonian vector field $\JJJ$, (ii) we have learned  that in order to make  the emergence of the pattern manifestly visible in solutions of the governing equations on the level $(\MMM,\JJJ)$,  we have to modify the Hamiltonian vector field $\JJJ$ by adding to it the generalized gradient term introduced in (\ref{eqgrr}). Note that without this modification (highlighting the recognised pattern) one can still in principle proceed but typically it becomes a daunting if not impossible task as in mathematical analysis of Landau damping\cite{villani2014}. An abstract time evolution equation in which the vector field is a sum of a Hamiltonian term (the right hand side of (\ref{Ham1}))   and the generalized gradient term (the right hand side of (\ref{eqgrr}) has been called GENERIC in \cite{go},\cite{og} (see more about the history  of this formulation for example in \cite{miroslav-guide,pkg}).

\subsection{GENERIC dynamics: $x\rightarrow y^{(fluid\, mech)}$ as $t\rightarrow \infty$}\label{3.4}

Our objective in this section is to show where does the fundamental flux-thermodynamic relation (\ref{eq0}) come from and how to choose  $K^{\dag}(y)$. As we were looking in the previous section for a vector field $J\in \mathfrak{X}(\MMM)$ that leads us to fixed points in $\MMM$, we are looking in this section for a vector field $\Upsilon\in \mathfrak{X}(\mathfrak{X}(\MMM))$ that will lead us also to fixed points but now the fixed points are  reduced vector fields $K\in\mathfrak{X}(\NNN)$. As an illustration,  we look for the vector field $\Upsilon$
in the particular setting in which  $\MMM$ is the state space of kinetic theory, $\JJJ \equiv \mathfrak{X}(\MMM)$ is the space of vector fields of kinetic theory, $\KKK\equiv \mathfrak{X}(\NNN)$ the space of vector fields of fluid mechanics, and $y=(\rho(\rr),\uu(\rr),e(\rr))\in \NNN$ are the state variables of fluid mechanics. Can we adapt the Boltzmann analysis presented above to this type of reduction?
Below, we make only a few steps in this direction. In particular, we shall not find the closure (we shall not find the fluid-mechanics constitutive relations) but we shall formulate it as a static and dynamic MaxEnt.

Our starting point is the Poisson-Grad hierarchy (\ref{eqs11}). We note that its analogue in the analysis of $x\rightarrow y^{(eq)}$  (see  the previous section)  is the set of equations
\begin{eqnarray}\label{BEhier}
\frac{d}{dt}\left(\begin{array}{cc}N\\E\end{array}\right)&=&\left(\begin{array}{cc}0\\0\end{array}\right)\nonumber \\
\frac{\partial f}{\partial t}&=&-\frac{\partial}{\partial r_i}\left(f\frac{\partial E^{\uparrow}(f)_f}{\partial v_i}\right)+\frac{\partial}{\partial v_i}\left(f\frac{\partial E^{\uparrow}(f)_f}{\partial r_i}\right)
\end{eqnarray}
and
\begin{equation}\label{BES}
\frac{dS}{dt}=0
\end{equation}
The second equation in (\ref{BEhier}) is the nondissipative  kinetic equation, the second equation in the Poisson-Grad hierarchy (\ref{eqs11}) is the nondissipative Poisson-Grad kinetic equation. This new kinetic equation  differs from the nondissipative  kinetic equation (\ref{kin1}) by the presence of  terms involving gradients of the hydrodynamic fields (the first two terms on the right hand side of the second equation in (\ref{eqs11})) and by the energy $E$ that in (\ref{eqs11})  depends  also on hydrodynamic fields. The reduction to equilibrium is made by investigating solutions to (\ref{kin1}), the reduction to fluid mechanics is made by investigating solutions to (\ref{kin1}).

As recalled in Section \ref{3.3},  a considerable amount of physical and mathematical insights collected in the last one hundred years about solutions of the Boltzmann kinetic equation allows us to say  (at least in the case when  $E^{\uparrow}(f)$ is only the kinetic energy) that there is a time independent pattern $\mathcal{P}^{(eq)}$ in the phase portrait corresponding to (\ref{BEhier}) and that this pattern  is revealed by following solutions to the Boltzmann kinetic equation (i.e. Eq.(\ref{kin1}) supplied with the Boltzmann collision term), or  still in a simple form, by following solutions to
\begin{equation}\label{Peq}
\frac{\partial f}{\partial t}=-\Lambda \Phi^{\uparrow}_f(f;E^*,N^*)
\end{equation}
to their conclusion; $\Lambda>0$ is a parameter and $\Phi^{\uparrow}(f;E^*,N^*)$ is the thermodynamic potential (\ref{eq2.21}) in which $x=f(\rr,\vv)$.

Due to the lack of physical and mathematical insights that would be comparable in their power to those collected for the Boltzmann equation (\ref{BEhier}), we limit ourselves in the investigation of solutions to the Poisson-Grad hierarchy only to a formal reformulation into dynamic and static MaxEnt principle.  The phase portrait  $\mathcal{P}^{(fl.mech.)}$ of fluid mechanics  emerges as a pattern in the phase portrait corresponding to
the MaxEnt reformulation of the Poisson-Grad hierarchy.
The statement that  the phase portrait  $\mathcal{P}^{(fl.mech.)}$ of fluid mechanics  emerges as a pattern in the phase portrait corresponding to the Poisson-Grad hierarchy  remains a conjecture.

The dynamic  MaxEnt reformulation of the Poisson-Grad kinetic equation that we are suggesting is a simple dynamical version of the reducing Legendre transformation discussed in Section \ref{2.2}:
\begin{equation}\label{PGeq}
\frac{\partial f^*}{\partial t}=-\Lambda \Psi^{\uparrow}_{f^*}(f;\mathbf{K}^{\dag})
\end{equation}
where $\Lambda>0$ is a parameter, $f^*=E_f$,
\begin{equation}\label{PsPG}
\Psi^{\uparrow}(f;\mathbf{K}^{\dag})
=-\mathfrak{S}^{\uparrow}(f^*)
+\int d\rr \int d\vv
\mathbf{K}^{\dag}\cdot \mathbf{K}(f^*)
\end{equation}
(see (\ref{eq01})),
\begin{eqnarray}\label{KK}
\mathbf{K}^{\uparrow}(f^*)&=&(K^{\uparrow (\rho)}(f^*),K^{\uparrow (u)}(f^*), K^{\uparrow (e)}(f^*))\nonumber \\
K^{\uparrow (\rho)}_i(f^*)&=&f\frac{\partial f^*}{\partial v_i}\nonumber \\
K^{\uparrow (u)}_{ij}(f^*)&=& f\left(f^*\delta_{ij}+v_i\frac{\partial f^*}{\partial v_j}\right)\nonumber \\
K^{\uparrow (e)}_i(f^*)&=& f\left(f^* E_{u_i}+\Pi\frac{\partial f^*}{\partial v_i}\right)
\end{eqnarray}
$\Pi=\rho E_{\rho}+\eta E_s+ff^*$.

The reduced fluxes $\mathbf{K}$  expressed in (\ref{KK}) as functions  of the distribution function
are read in the second term on the right hand side of the first equation in the Poisson-Grad hierarchy (\ref{eqs11}). On the other hand, in order to specify  the flux-entropy $\mathfrak{S}^{\uparrow}(f^*)$ as well as the specification of  the Lagrange multipliers $\mathbf{K}^{\dag}$ we have to begin to investigate  trajectories generated by the right hand side of (\ref{eqs11}). Indeed, we recall that the Boltzmann entropy (that plays the role of  $\mathfrak{S}^{\uparrow}(f^*)$ in the investigation of the reduction to equilibrium) is not directly seen in the vector field.
As for the Lagrange multipliers $\mathbf{K}^{\dag}$,  we  can  read their basic form in the first two terms on the right hand side of the Poisson-Grad kinetic equation (i.e. the second equation in the Poisson-Grad hierarchy).

The time evolution governed by the gradient dynamics (\ref{PGeq}) is clearly making the reducing Legendre transformation discussed in Section \ref{2.2}. Consequently, the above reformulation of the Poisson-Grad hierarchy introduces a fundamental flux-thermodynamic relation to the level of fluid mechanics.

We end this section with a simple illustration in which some additional simplifications and additional physical arguments make the above formal reformulation of the Poisson-Grad hierarchy more explicit.
Being inspired by the  Boltzmann strategy in the context of the investigation of the passage  $J\rightarrow K$, we ask the question of what could be  the principal source of  complexity of solutions to the kinetic equation that allows to introduce a regularizing dissipative term (that, in the Boltzmann equation is the Boltzmann collision term)  simplifying the solutions.
Following  Boltzmann,  we  suggest that the irregularities in solutions arise in the momentum variable $\vv$. A  microscopic turbulence emerges. This insight into the importance of the dependence on the momentum we then express mathematically by suggesting that the Fokker-Planck term $\frac{\partial}{\partial v_i}\left(f\Lambda \frac{\partial E_f}{\partial v_i}\right)$, where $\Lambda>0$ is a parameter,  could be  the regularizing dissipative term added to the Poisson-Grad kinetic equation. In order to keep the equation governing the time evolution of the energy field $e(\rr)$ unchanged, we modify also the equation (\ref{seq}) by adding to its right hand side the entropy production

Now being inspired by the Chapman-Enskog analysis of solutions of the Boltzmann equation, we look for  dominant terms on the right side of the Poisson-Grad  kinetic equation. One such term will be a dissipative term
but the terms in which the coupling to the hydrodynamic fields is expressed, i.e. the terms the extra fluxes (the terms in the first line on the right hand side of the PG equation) and the extra forces (the terms in the second line on the right hand side of the PG equation), are also important. Having  in mind our  anticipation  of the microscopic turbulence, we assume  that the extra forces in the PG equation will play more important role. Finally, we assume that the term involving the gradient of the hydrodynamic momentum is more important than the terms involving gradients the remaining hydrodynamic fields. Consequently, the zero Chapman-Enskog approximation of the PG kinetic equation is
\begin{equation}\label{CReq}
\frac{\partial f}{\partial t}=-\frac{\partial}{\partial v_i}\left(f v_j\frac{\partial u^*_j}{\partial r_i}\right)+\frac{\partial}{\partial v_i}\left(f\Lambda \frac{\partial f^*}{\partial v_i}\right)
\end{equation}
This equation can also written as
\begin{equation}\label{end1}
\frac{\partial f}{\partial t}=  \Psi^{\uparrow}_{f^*}
\end{equation}
where $\Psi^{\uparrow}_{f^*}=-\mathfrak{S}^{\uparrow}(f^*)+\int d\rr \int d\vv \left(fv_j\frac{\partial E_{u_j}}{\partial r_i}\right)\frac{\partial f^*}{\partial v_i}$, and $\mathfrak{S}^{\uparrow}(f^*)=\frac{1}{2}\int d\rr \int d\vv \Lambda f\frac{\partial f^*}{\partial v_i} \frac{\partial f^*}{\partial v_i}$.

Equation $\Psi^{\uparrow}_{f^*}=0$ implies $\Lambda f \frac{\partial f^*}{\partial v_i}=f v_k\frac{\partial E_{u_k}}{\partial r_i}$.  By multiplying this equation by $v_j$ and integrating it over $\vv$ we obtain
\begin{equation}\label{end2}
\int d\vv f v_j \frac{\partial f^*}{\partial v_i}=
\frac{\Gamma}{\Lambda}\frac{\partial E_{u_j}}{\partial r_i}
\end{equation}
if we assume that $\int d\vv f v_i v_j=\Gamma \delta_{ij}$, $\Gamma>0$ is a parameter. The left hand side of this equation is the stress tensor (see the third  equation in (\ref{KK})) and the right hand side is the Navier-Stokes constitutive relation for the stress tensor, $\frac{\Gamma}{2\Lambda} $ is the viscosity coefficient.

\subsection{Reduction to hydrodynamic fields}
%\todo{vk: prijde mi clanek napsany opravdu pekne, srozumitelne, ale me osobne by hodne pridala ilustrace ci spise pouziti teto komplexni obecne metody na neco noveho. Zkratka mit tu ``punch line'' pro recenzenty. V tomto ohledu mi neprisel spatny dodatek Michala, ktery to drobne upraveny znovu prikladam. Neni to vsak jaksi hotove, jeste na tom dal pracuji...}

%\todo{jaka je role toho pridavaneho regularizing clenu? Lze napsat souvis s 1/eps faktorem? Nemame ani predtim, ale bylo by to pekne}
%\todo{vse prepsat jako poravdu konjugaty vzhl entropii, jak o tom mluvime v hlavni casti textu; dava smysl napsat to pak i konkretne pro id plyn?}

We repeat the pattern recognition argument from the previous section and let the distribution function relax to the fixed point. As a result, we obtain a variant of description of hydrodynamics that includes a higher degree of microscopic effects (micro turbulence in velocity).

We again add the regularizing dissipative (Fokker-Planck like) term to the evolution equation for distribution function and assume as done above that its effect is time-scale separation among the terms. We expect the dominant balance between time derivative, the contribution from the velocity flux $J^{v_i}$ and the regularizing dissipative term. %\todo{VK: jsem ted trochu zase zmaten, co je vhodna volba pro dominantni cleny. V casti vyse vybirame cleny s hydrodyn rychlosti jakozto velike, ale vynechavame napr $\nabla(f E_{u_i})$ a tez neni mi jasne, proc $\partial_t f$ ma byt tez dulezity.. Je to trochu jedno, at vybereme skoro cokoliv, jde to takto napsat, jen po projekci, dosazeni fixed pointu, dostanem neco trochu jineho. A to co dostavame nyni se zda byti dobre..}
More precisely, we assume the following structure of the evolution equation for the distribution function:
\begin{subequations}\label{eqs.reg}
\begin{eqnarray}
  \frac{\partial f}{\partial t}%&=& -\frac{\partial}{\partial r_i}\left[f\left(E_{u_i}+\frac{\partial\eta_f}{\partial v_i}E_s\right)\right] \nonumber \\
%&&+\frac{\partial}{\partial v_i}\left[f\left(\frac{\partial E_{\rho}}{\partial r_i}+\frac{\partial (\eta_fE_s)}{\partial r_i}+v_j\frac{\partial E_{u_j}}{\partial r_i}\right)\right]\nonumber \\
%&&-\frac{\partial}{\partial r_i}\left(f\frac{\partial E_f}{\partial v_i}\right)+\frac{\partial}{\partial v_i}\left(f\frac{\partial E_f}{\partial r_i}\right)
%   +\frac{\partial}{\partial v_i} \left(f \Lambda_{ij} \frac{\partial E_f}{\partial v_j}\right)\\
 &=& -\epsilon\frac{\partial}{\partial r_i}\left[f\left(E_{u_i}+\frac{\partial\eta_f}{\partial v_i}E_s+\frac{\partial E_f}{\partial v_i}\right)\right] \\
&&+\frac{\partial}{\partial v_i}\left[f\left(\frac{\partial E_{\rho}}{\partial r_i}+\frac{\partial (\eta_fE_s)}{\partial r_i}+v_j\frac{\partial E_{u_j}}{\partial r_i}+\frac{\partial E_f}{\partial r_i}\right)\right]\nonumber \\
&& +\frac{\partial}{\partial v_i} \left(f \Lambda_{ij} \frac{\partial E_f}{\partial v_j}\right),\nonumber
\end{eqnarray}
with $\epsilon$ being a small parameter and where we uniquely identified the ``spatial and velocity fluxes'', $J^{r_i}$ and $J^{v_i}$, as terms in divergences w.r.t $r_i$ and $v_i$.. The hydrodynamic fields are unaffected explicitly by the scaling yielding the governing equations of the modified Poisson-Grad hierarchy
\begin{eqnarray}
\frac{\partial\rho}{\partial t} &=& -\frac{\partial (\rho E_{u_i})}{\partial r_i}
	-\frac{\partial}{\partial r_i}\int d\vv  f\frac{\partial E_f}{\partial v_i}\\
\frac{\partial u_i}{
\partial t} &=& -\frac{\partial (u_iE_{u_j})}{\partial r_j}-\frac{\partial p}{\partial r_i}
	-\frac{\partial}{\partial r_j} \int d\vv f v_i\frac{\partial E_f}{\partial v_j}\\
\frac{\partial s}{\partial t} &=& -\frac{\partial (s E_{u_i})}{\partial r_i}
	-\frac{\partial}{\partial r_i}\int d\vv \eta \frac{\partial E_f}{\partial v_i}
	+\frac{1}{E_s}\int d\vv f \Lambda_{ij} \frac{\partial E_f}{\partial v_i}\frac{\partial E_f}{\partial v_j}.
\end{eqnarray}
\end{subequations}
Equations \eqref{eqs.reg} consist of reversible (Hamiltonian) and irreversible part. The irreversible part is represented by the Fokker-Planck-like dissipation (the last term in the equation for $f$) and the corresponding entropy production. Using the transformation between the conjugate variables in the energetic representation (derivatives of energy) and entropic representation (derivatives of entropy, denoted by stars), we can write
\begin{equation}
	E_f = - \frac{S_f}{S_e} = - \frac{f^*}{e^*}.
\end{equation}
See \cite{shtc-generic,pkg,callen} for more details. Note that no dissipative evolution appears in the equation for the total spatial energy density. The Fokker-Planck-like dissipation can be seen as derivative of dissipation potential
\begin{equation}
	\Xi(f^*) = \frac{1}{2}\int d\rr\int d\vv \Lambda T f\frac{\partial f^*}{\partial v_i}\frac{\partial f^*}{\partial v_i}
\end{equation}
with respect to $f^*$, where $T=\frac{\partial e}{\partial s}$ stands for the local temperature field. The irreversible Fokker-Planck-like terms can be motivated in two ways: (i) Considering a continuum particle is subject to random (Brownian) motion, the Fokker-Planck term appears in the equation for the distribution function as a result of  the fluctuations \cite{hco}, and (ii) it is anticipated that fast oscillations in the $\vv-$space develop due to phenomena related to the Landau damping \cite{villani2014,krm,pavelka-ld,despres-damping}.

Focusing on the leading order solution of the distribution function, we take advantage of the fact that its evolution is described by a first order linear partial differential equation whose characteristics, parametrized by time $t$, are given by
\begin{equation*}
  \frac{d}{d t} R_i=\epsilon J^{r_i}_f,\quad \frac{d}{d t} V_i = J^{v_i}_f.
\end{equation*}
Note that both fluxes $J^{r_i}$, $J^{v_i}$ are linear in $f$ hence $J^{r_i}_f=J^{r_i}/f$ and similarly the other one. 
The leading order solution to this problem is also known as the inner solution in the singular perturbation method, while rescaling of time $t=\tau\epsilon$ provides the outer problem
\begin{equation*}
  \frac{d}{d t} R_i= J^{r_i}_f,\quad \epsilon \frac{d}{d t} V_i = J^{v_i}_f.
\end{equation*}
yielding the large-time (outer) solution. Note that for the fast initial period yields a fast equilibration of $V_i(t)$ takes place while characteristics remain almost parallel with $v_i$ ($R_i(t)$ are essentially constant); for larger times (the outer solution) $V_i$ are enslaved to the remaining state variables (invoking singular perturbation limit and an analogue of Tikhonov's theorem).
%\todo{MP: Nevim, jestli to chapu spravne, ale nemely by charakteristiky byt videt az z rovnice typu $\partial_t f + \dot{R}^i \frac{\partial f}{\partial r^i} + \dot{V}_i \frac{\partial f}{\partial v_i}$, pricemz $(\dot{R}^i,\dot{V}_i)$ je tecny vektor k charakteristice?}

Hence here we can see a particular realisation of the convergence of the upper vector fields to the lower level vector fields corresponding to large time asymptotics of the characteristics, with a fixed point being a solution to
\begin{equation} \label{eq.ConstitRel}
  0=\frac{\partial}{\partial v_i} J^{v_i}.
\end{equation}

Before proceeding further, we note explicitly the relation to the general approach outlined above.
The evolution of distribution function at the leading order can also written as
\begin{equation*}
\frac{\partial f}{\partial t}=  \Psi^{\uparrow}_{f^*}
\end{equation*}
where $$\Psi^{\uparrow}_{f^*}=-\mathfrak{S}^{\uparrow}(f^*)+\int d\rr \int d\vv \left(f\left[\frac{\partial E_{\rho}}{\partial r_i}+\frac{\partial (\eta_f E_{s})}{\partial r_i}+v_j\frac{\partial E_{u_j}}{\partial r_i}+\frac{\partial E_f}{\partial r_i}\right]\right)\frac{\partial f^*}{\partial v_i},$$ and $\mathfrak{S}^{\uparrow}(f^*)=\frac{1}{2}\int d\rr \int d\vv \Lambda f\frac{\partial f^*}{\partial v_i} \frac{\partial f^*}{\partial v_i}$.

The fixed point given in \eqref{eq.ConstitRel} corresponds to equation $\Psi^{\uparrow}_{f^*}=0$ and implies
\begin{equation}\label{eq.const}
  \frac{\partial E_{\rho}}{\partial r_i}+\frac{\partial (\eta_fE_s)}{\partial r_i}+v_j\frac{\partial E_{u_j}}{\partial r_i} +\frac{\partial E_f}{\partial r_i}
= - \Lambda_{ij} \frac{\partial E_f}{\partial v_j}.
\end{equation}
One should view this condition as a relation yielding the leading order distribution function as a solution rather than a constitutive relation for $E_f$. Finally, we may project the extended hydrodynamic fields $\rho,\uu,s,f$ to hydrodynamic fields $\rho,\uu,s$ via MaxEnt in the vector fields as described above. In this particular case we may simply plug the eq. \eqref{eq.const} back into the regularized Eqs. \eqref{eqs.reg} to obtain the reduced system
\begin{subequations}
\begin{eqnarray}\label{eqs.relax}
\frac{\partial\rho}{\partial t}
 &=& -\frac{\partial (\rho E_{u_i})}{\partial r_i}\\
	&&+\frac{\partial}{\partial r_i}
	\left(
	\frac{\partial E_{\rho}}{\partial r_i}\int d\vv \frac{f}{\Lambda}
	+\int d\vv \frac{f}{\Lambda} \frac{\partial (\eta_f E_s)}{\partial r_i}
	+\frac{\partial E_{u_j}}{\partial r_i}\int d\vv v_j \frac{f}{\Lambda}
	+\int d\vv \frac{f}{\Lambda} \frac{\partial E_f}{\partial r_i} \right)\nonumber
	\\
\frac{\partial u_i}{\partial t} &=& -\frac{\partial (u_iE_{u_j})}{\partial r_j}-\frac{\partial p}{\partial r_i}\\
	&&+\frac{\partial}{\partial r_j} \left(
	+\frac{\partial E_{\rho}}{\partial r_j}\int d\vv \frac{f v_i}{\Lambda}
	+\int d\vv \frac{f v_i}{\Lambda}\frac{\partial (\eta_fE_s)}{\partial r_j}
	+\frac{\partial E_{u_k}}{\partial r_j}\int d\vv \frac{f v_i v_k}{\Lambda}
	+\int d\vv \frac{f v_i}{\Lambda} \frac{\partial E_f}{\partial r_j} \right)\nonumber\\
	\partial_t s &=& -\frac{\partial (s E_{u_i})}{\partial r_i} - \partial_i J^{(s)}_i + \sigma_s
\end{eqnarray}
where
\begin{eqnarray}
	J^{(s)}_i &=& -\int d\vv \eta \frac{
		\frac{\partial E_{\rho}}{\partial r_i}+\frac{\partial (\eta_fE_s)}{\partial r_i}+v_j\frac{\partial E_{u_j}}{\partial r_i} +\frac{\partial E_f}{\partial r_i}}{\Lambda}\\
	\sigma_s &=&\frac{1}{E_s}\int d\pp \frac{f}{\Lambda}\left(
	\frac{\partial E_{\rho}}{\partial r_i}+\frac{\partial (\eta_f E_s)}{\partial r_i}+v_j\frac{\partial E_{u_j}}{\partial r_i} +\frac{\partial E_f}{\partial r_i} \right)^2.
\end{eqnarray}
\end{subequations}
%These equation can be seen as hydrodynamic equation with dissipation being divergence of dissipative fluxes, each flux being of form
%\begin{equation}
%	J^{x} = M^{xy} X_{y}
%\end{equation}
%where $\mathbf{M}$ is a matrix and $\mathbf{X} = (\nabla E_\rho, \nabla E_\uu, \nabla E_s)$ is the vector of thermodynamic forces. The hydrodynamic equations are then equipped with a relaxed equation for the distribution function, which is advected by the hydrodynamic fields.
Neglecting the off-diagonal terms of the matrix, i.e. assuming for simplicity a case without any non-trivial coupling (trivial corresponds to advection of fields), the relaxed equations become
\begin{subequations}
\begin{eqnarray}\label{eqs.relax.diag}
\frac{\partial\rho}{\partial t} &=& -\frac{\partial (\rho E_{u_i})}{\partial r_i}
	+\frac{\partial}{\partial r_i}
	\left(
	\frac{\partial E_{\rho}}{\partial r_i}\int d\vv \frac{f}{\Lambda}\right)\\
\frac{\partial u_i}{\partial t} &=& -\frac{\partial (u_iE_{u_j})}{\partial r_j}-\frac{\partial p}{\partial r_i}
	+\frac{\partial}{\partial r_j} \left(
	\frac{\partial E_{u_k}}{\partial r_j}\int d\vv \frac{f v_i v_k}{\Lambda}\right)\\
\frac{\partial s}{\partial t} &=& -\frac{\partial (s E_{u_i})}{\partial r_i} - \partial_i J^{(s)}_i + \sigma_s\\
\frac{\partial f}{\partial t}&=& -\frac{\partial}{\partial r_i}\left[f\left(E_{u_i}+\frac{\partial\eta_f}{\partial v_i}E_s\right)\right]
	+\frac{\partial}{\partial r_i}\left(\frac{f}{\Lambda}\frac{\partial E_f}{\partial r_i}\right),
\end{eqnarray}
which can be seen as hydrodynamic equations with self-diffusion similar to \cite{svard}.%\todo{doplnit komentar k ziskanym rovnicim - dokonce jsou rozumne, asi..}
In \cite{svard} a kinetic theory with explicit diffusion in the $\rr-$space was proposed, and as a result of the projection to the hydrodynamic fields, Laplacians appear on the right hand sides of the equations for density, momentum density and energy density. Such an alternative to the Navier-Stokes equations seem to be advantageous from both the mathematical and numerical points of view.
\end{subequations}

To close the equations we need to specify the microscopic entropy, $\eta(f)$, and substitute the leading order solution $f$ to \eqref{eq.const}. For example, in the case of ideal gas we may combine the local Sackur-Tetrode equation of state (its inverse to obtain $e(s)$) for hydrodynamic fields which is obtained by a projection from the kinetic theory with a one particle distribution function and energy containing just the kinetic energy \cite{pkg}. To follow the idea of extending the energy by a distribution function dependence, we suggest to combine these two energies to have
\begin{equation*}
E = \int d \rr \left[\frac{1}{2}\left(\frac{\uu^2}{2 \rho} + \frac{3 h^2}{4 \pi m} \left[\frac{\rho}{m}\right]^{5/3}\exp\left[\frac{2}{3}\left(\frac{m s}{k_B\rho}-\frac{5}{2}\right)\right]\right) +\int d \vv \frac{1}{2} \frac{\vv^2}{2m}f\right]
\end{equation*}
In such a case, the only term in the relation for the leading order distribution function is  $\eta= -k_B f (\ln(h^3f)-1)$ and allows explicit form of solution in terms of the hydrodynamic fields. The motivation for such choice of energy can be seen in the grand-canonical BBGKY hierarchy \cite{hierarchy}, where energy is expressed as the sum of energies on different levels of description. Using this energy and entropy $\eta(f)$ and assuming that $\Lambda = \mathrm{const}$, the equation for density becomes
\begin{subequations}
\begin{equation}
	\partial_t \rho = -\partial_j(u_j) + 
	\frac{\partial}{\partial r_i} \left(\frac{\rho}{\Lambda}\frac{\partial \mu}{\partial r_i}-\frac{k_B}{\Lambda} \frac{\partial \rho T}{\partial r_i} +\frac{1}{\Lambda} s \frac{\partial T}{\partial r_i}\right),
\end{equation}
	where the diffusive term involving gradient of the chemical potential $\mu$ is revealed explicitly. Note the explicit presence of extra mass flux (density not only being advected), which was advocated in \cite{Bre06a,Bre12a} and \cite{BedEta06a,Miroslav-mass-flux}, opposed in \cite{HCO-mass-flux} and brought up again in \cite{extra-mass-flux}, where an example satisfying all criteria from \cite{HCO-mass-flux} was constructed while still having an extra mass flux. We consider the discussion still open. Assuming the local-equilibrium distribution function, the term in the equation for momentum density dependent on $f$ becomes $\nu \delta_{ik}$, $\nu$ being a viscosity coefficient, and the Navier-Stokes dissipation appears, 
\begin{equation}
\frac{\partial u_i}{\partial t} = -\frac{\partial (u_iE_{u_j})}{\partial r_j}-\frac{\partial p}{\partial r_i}
	+\frac{\partial}{\partial r_j} \left(
	\nu \frac{\partial E_{u_i}}{\partial r_j}\right).
\end{equation}
The evolution equation for entropy density contains irreversible terms expressing heat conduction and entropy production.
\end{subequations}
Another properties of the reduced equations, which represent a new version of the Chapman-Enskog expansion \cite{dgm}, are left for future research.

\section{Discussion}\label{4}

There are two main results in this paper. First, it is a unified formulation of reductions among mesoscopic theories (both without and with the time evolution)  of macroscopic systems, and second, it is the  Poisson-Grad hierarchy.

An autonomous mesoscopic model of macroscopic systems is always enriched by relating it to more microscopic models.
In  reductions to equilibrium  models, the gain is thermodynamics (more precisely the  fundamental thermodynamic relation). It represents an information inherited from the way the  equilibrium model has emerged as a pattern in solutions of the governing equations in the more detailed theory. In  reductions to less detailed mesoscopic dynamical models  the  gain is the reduced dynamics but also an additional information inherited from the way the less detailed dynamics emerged as a pattern in solutions to the more detailed dynamics. In analogy with the reduction to the equilibrium theory, we  call  this new addition (now  to a mesoscopic dynamical theory)  a flux-thermodynamics (more precisely a  fundamental flux-thermodynamic relation).

The Poisson-Grad hierarchy is a new reformulation of general kinetic equations that couples kinetic theory with hydrodynamics while preserving the Hamiltonian kinematics of both theories. In this reformulation, the one particle distribution function represents an extra microscopic information that is invisible in continuum mechanics. Except for a few observations made in the last section of this paper, where an alternative to the Chapman-Enskog expansion is proposed, the problem of investigating solutions to the Poisson-Grad hierarchy remains an open problem.

%\ethics{Insert ethics statement here if applicable.}

%\dataccess{Insert details of how to access any supporting data here.}

%\aucontribute{MG developed the main ideas and wrote most of the manuscript. VK and MP added concrete illustrations and finalized the manuscript.}

%\competing{The author(s) declare that they have no competing interests.}

%\funding{M.G. was supported by the Natural Sciences and Engineering Research Council of Canada, Grants 3100319 and 3100735.

\section*{Acknowledgment}
M.P. and V.K. were supported by Czech Science Foundation, Project No. 17-15498Y, and by Charles University Research Program No. UNCE/\-SCI/023.
M.P. is grateful to Magnus Sv{\" a}rd for discussions on extra mass flux and alternatives to Navier-Stokes. Authors are grateful to P{\' e}ter V{\' a}n for his kind invite into this special issue.

%\bibliographystyle{vancouver}
%\bibliography{library}

\begin{thebibliography}{10}

\bibitem{jaynes}
Jaynes ET.
\newblock Foundations of probability theory and statistical mechanics.
\newblock In: Delaware Seminar in the Foundation of Physics, M. Bunge ed.
  Springer New York; 1967. .

\bibitem{pkg}
Pavelka M, Klika V, Grmela M.
\newblock Multiscale Thermo-Dynamics.
\newblock de Gruyter (Berlin); 2018.

\bibitem{jizba-maxent}
Jizba P, Korbel J.
\newblock Maximum Entropy Principle in Statistical Inference: Case for
  Non-Shannonian Entropies.
\newblock Phys Rev Lett. 2019 Mar;122:120601.
\newblock Available from:
  \url{https://link.aps.org/doi/10.1103/PhysRevLett.122.120601}.

\bibitem{redext}
Grmela M, Klika V, Pavelka M.
\newblock Reductions and extensions in mesoscopic dynamics.
\newblock Phys Rev E. 2015;92(032111).

\bibitem{dynmaxent}
Klika V, Pavelka M, V{\' a}gner P, Grmela M.
\newblock Dynamic Maximum Entropy Reduction.
\newblock Entropy. 2019;21(715).

\bibitem{boltzmann}
Gesamtausgabe LB.
\newblock Ludwig Boltzmann Gesamtausgabe - Collected Works; 1983.

\bibitem{gibbscw}
Gibbs JW.
\newblock Collected Works.
\newblock Longmans; Green and Comp. New York; 1984.

\bibitem{grmela2019entropy}
Grmela M, Pavelka M, Klika V, Cao BY, Bendian N.
\newblock Entropy and entropy production in multiscale dynamics.
\newblock Journal of Non-Equilibrium Thermodynamics. 2019;.

\bibitem{marsden-bbgky}
Marsden JE, Morrison PJ, Weinstein A.
\newblock The hamiltonian structure of the BBGKY hierarchy equations.
\newblock Cont Math AMS. 1984;28:115--124.

\bibitem{grad}
Grad H.
\newblock Principles of Kinetic Theory of Gases.
\newblock In: Encyclopedia of Physics. vol.~12. Springer-Verlag; 1958. .

\bibitem{miroslav-grad}
Grmela M, Hong L, Jou D, Lebon G, Pavelka M.
\newblock Hamiltonian and Godunov structures of the Grad hierarchy.
\newblock Physical Review E. 2017;95(033121).

\bibitem{rugjap}
Ruggeri T, Sugiyama M.
\newblock Rational Extended Thermodynamics Beyond the Monoatomic Gas.
\newblock Springer, Heidelberg; 2015.

\bibitem{dgm}
de~Groot SR, Mazur P.
\newblock Non-equilibrium Thermodynamics.
\newblock New York: Dover Publications; 1984.

\bibitem{van-berezovski}
Berezovski A, V{\'a}n P.
\newblock Internal Variables in Thermoelasticity.
\newblock Solid Mechanics and Its Applications. Springer International
  Publishing; 2017.

\bibitem{ch}
Cahn JW, Hilliard JE.
\newblock Free Energy of a Nonuniform System. {I}nterfacial Free Energy.
\newblock Journal of Chemical Physics. 1958;28(258).

\bibitem{landau-ginzburg}
Ginzburg VL, Landau LD.
\newblock On the theory of superconductivity.
\newblock Zhur Eksp Theor Fiz. 1950;20:1064--1082.

\bibitem{clebsch}
Clebsch A.
\newblock \"{U}ber die {I}ntegration der {H}ydrodynamische {G}leichungen.
\newblock Journal f\"{u}r die reine und angewandte Mathematik. 1895;56:1--10.

\bibitem{arnold}
Arnold VI.
\newblock Sur la g\'{e}ometrie diff\'{e}rentielle des groupes de Lie de
  dimension infini et ses applications dans l'hydrodynamique des fluides
  parfaits.
\newblock Annales de l'institut Fourier. 1966;16(1):319--361.

\bibitem{ehrenfests}
Ehrenfest P, Ehrenfest T.
\newblock The Conceptual Foundations of the Statistical Approach in Mechanics.
\newblock Dover Books on Physics. Dover Publications; 1990.

\bibitem{gk-ehrenfest}
Gorban AN, Karlin IV, \"{O}ttinger HC, Tatarinova LL.
\newblock Ehrenfest’s argument extended to a formalism of nonequilibrium
  thermodynamics.
\newblock Physical Review E. 2001;63(066124).

\bibitem{ehrenfest-regularization}
Pavelka M, Klika V, Grmela M.
\newblock Ehrenfest regularization of Hamiltonian systems.
\newblock Physica D: Nonlinear Phenomena. 2019;399:193 -- 210.
\newblock Available from:
  \url{http://www.sciencedirect.com/science/article/pii/S0167278918305232}.

\bibitem{villani2014}
Villani C.
\newblock {Particle systems and nonlinear Landau damping}.
\newblock {Physics of plasmas}. {2014} {MAR};{21}({3}).

\bibitem{go}
Grmela M, \"{O}ttinger HC.
\newblock Dynamics and thermodynamics of complex fluids. {I}. {D}evelopment of
  a general formalism.
\newblock Phys Rev E. 1997 Dec;56:6620--6632.
\newblock Available from:
  \url{http://link.aps.org/doi/10.1103/PhysRevE.56.6620}.

\bibitem{og}
\"Ottinger HC, Grmela M.
\newblock Dynamics and thermodynamics of complex fluids. {II}. {I}llustrations
  of a general formalism.
\newblock Phys Rev E. 1997 Dec;56:6633--6655.

\bibitem{miroslav-guide}
Grmela M.
\newblock GENERIC guide to the multiscale dynamics and thermodynamics.
\newblock J Phys Commun. 2018;2(032001).

\bibitem{shtc-generic}
Peshkov I, Pavelka M, Romenski E, Grmela M.
\newblock Continuum Mechanics and Thermodynamics in the {H}amilton and the
  {G}odunov-type Formulations.
\newblock Continuum Mechanics and Thermodynamics. 2018;30(6):1343--1378.

\bibitem{callen}
Callen HB.
\newblock Thermodynamics: an introduction to the physical theories of
  equilibrium thermostatics and irreversible thermodynamics.
\newblock Wiley; 1960.
\newblock Available from: \url{http://books.google.cz/books?id=mf5QAAAAMAAJ}.

\bibitem{hco}
{\"O}ttinger HC.
\newblock Beyond Equilibrium Thermodynamics.
\newblock Wiley; 2005.

\bibitem{krm}
Grmela M, Pavelka M.
\newblock Landau damping in the multiscale {V}lasov theory.
\newblock Kinetic and Related Models. 2018;11(3):521--545.

\bibitem{pavelka-ld}
Pavelka M, Klika V, Grmela M.
\newblock Thermodynamic explanation of Landau damping by reduction to
  hydrodynamics.
\newblock Entropy. 2018;20.

\bibitem{despres-damping}
Despr{\' e}s B.
\newblock Scattering Structure and Landau Damping for Linearized {V}lasov
  Equations with Inhomogeneous {B}oltzmannian States.
\newblock Ann Henri Poincar{\' e}. 2019;20:2767–2818.

\bibitem{svard}
Sv{\" a}rd M.
\newblock A new Eulerian model for viscous and heat conducting compressible
  flows.
\newblock Physica A: Statistical Mechanics and its Applications. 2018;506:350
  -- 375.

\bibitem{hierarchy}
Pavelka M, Klika V, Esen O, Grmela M.
\newblock A hierarchy of {P}oisson brackets in non-equilibrium thermodynamics.
\newblock Physica D: Nonlinear phenomena. 2016;335:54--69.

\bibitem{Bre06a}
Brenner H.
\newblock Fluid mechanics revisited.
\newblock Physica A: Statistical Mechanics and its Applications.
  2006;370(2):190--224.

\bibitem{Bre12a}
Brenner H.
\newblock Beyond {N}avier--{S}tokes.
\newblock International Journal of Engineering Science. 2012;54:67--98.

\bibitem{BedEta06a}
Bedeaux D, Kjelstrup S, {\"O}ttinger HC.
\newblock On a possible difference between the barycentric velocity and the
  velocity that gives translational momentum in fluids.
\newblock Physica A: Statistical Mechanics and its Applications.
  2006;371(2):177--187.

\bibitem{Miroslav-mass-flux}
Grmela M.
\newblock Mass flux in extended and classical hydrodynamics.
\newblock Physical Review E. 2014;89(063024).

\bibitem{HCO-mass-flux}
\"{O}ttinger HC, Struchtrup H, Liu M.
\newblock Inconsistency of a dissipative contribution to the mass flux in
  hydrodynamics.
\newblock Physical Review E. 2009;80(5):056303.

\bibitem{extra-mass-flux}
V{\' a}n P, Pavelka M, Grmela M.
\newblock Extra Mass Flux in Fluid Mechanics.
\newblock Journal of Non-Equilibrium Thermodynamics. 2016;42(2):133--151.

\end{thebibliography}

\end{document}